\documentclass[a4paper]{article}

\usepackage[utf8]{inputenc}
\usepackage{graphicx}

\usepackage[english]{babel}
\usepackage{amsmath}
\usepackage{amssymb}
\usepackage{amsthm}
\usepackage{dsfont}
\usepackage{cases}
\usepackage[shortlabels]{enumitem}
\usepackage{color}
\usepackage{hyperref}
\usepackage{verbatim}
\usepackage{setspace}

\usepackage{geometry}

\newcommand*{\email}[1]{\href{mailto:#1}{\nolinkurl{#1}}} 

\usepackage{tikz}
\definecolor{darkgreen}{rgb}{0.1,.8,0.1}
\definecolor{darkred}{rgb}{0.8,.1,0.1}
\definecolor{darkblue}{rgb}{0.1,.1,0.8}

\usepackage{pgfplots}
\pgfplotsset{compat=newest}

\theoremstyle{plain}
\newtheorem{thm}{Theorem}[section]
\newtheorem{lem}[thm]{Lemma}

\theoremstyle{definition}
\newtheorem{defn}{Definition}[section]

\newtheorem{ass}{Assumption}[section]

\theoremstyle{remark}

\def \E{\mathbb{E}}

\def \P{\mathbb{P}}

\def \R{\mathbb{R}}

\def\erm{\mathrm{e}}
\def\drm{\mathrm{d}}

\def \Tc{\mathcal{T}}

\def \Bc{\mathcal{B}}
\def \Dc{\mathcal{D}}

\def \Ck{\mathfrak{C}}
\def\bmin{{\beta_{min}}}

\DeclareMathOperator*{\argmin}{arg\,min}

\begin{document}

\title{Contact rate epidemic control of COVID-19: an equilibrium view}

\author{Romuald {\sc Elie}\footnote{LAMA - UMR 8050, Universit\'e Gustave Eiffel, France; \email{romuald.elie@univ-mlv.fr}}
\and Emma {\sc Hubert}\footnote{LAMA - UMR 8050, Universit\'e Gustave Eiffel, France; \email{emma.hubert@univ-paris-est.fr}}
\and Gabriel {\sc Turinici}\footnote{CEREMADE - UMR 7534, Université Paris Dauphine, PSL Research University, France, \email{gabriel.turinici@dauphine.fr}}}

\date{\today} 
\maketitle

\begin{abstract}
We consider the control of the COVID-19 pandemic through a standard SIR compartmental model. This control is induced by the aggregation of individuals' decisions to limit their social interactions: when the epidemic is ongoing, an individual can diminish his/her contact rate in order to avoid getting infected, but this effort comes at a social cost. If each individual lowers his/her contact rate, the epidemic vanishes faster, but the effort cost may be high. A Mean Field Nash equilibrium at the population level is formed, resulting in a lower effective transmission rate of the virus. We prove theoretically that equilibrium exists and compute it numerically. However, this equilibrium selects a sub-optimal solution in comparison to the societal optimum (a centralized decision respected fully by all individuals), meaning that the cost of anarchy is strictly positive. We provide numerical examples and a sensitivity analysis, as well as an extension to a SEIR compartmental model to account for the relatively long latent phase of the COVID-19 disease. In all the scenarii considered, the divergence between the individual and societal strategies happens both before the peak of the epidemic, due to individuals' fears, and after, when a significant propagation is still underway.

\medskip

\noindent
\textbf{Mathematics Subject Classification.} 92D30, 92Bxx, 91A16

\medskip

\noindent
\textbf{Keywords.} COVID-19, SARS-CoV-2, Epidemic control, SIR model, Mean Field Games

\end{abstract}

\section*{Introduction}

As of April 10th 2020, almost half of world population is under strong restrictions (lockdown) enforced by local governments to limit the ongoing SARS-CoV-2 mediated COVID-19 pandemic. These restrictions mainly concern the reduction of social interactions. They are instantiated at different geographical scales (house-holdings, towns, states, countries...) and already had a significant worldwide economic impact. All individuals affected by epidemic control measures pay a cost in terms of health (reports show an under-reporting of severe illnesses such as hearth, ischemic strokes, ...), time, money, social interactions, psychological pressure (domestic violence increased), etc. However, until a sufficient proportion of the population becomes immune (by infection or vaccination), the choice of control measures together with an adequate tracking of the virus spread are key to limit the epidemic as well as the social and economic impacts of the contact rate decay. In such context, each individual needs to balance the individual and collective outcomes of his/her behavior, when choosing his/her level of social interaction with the population. For instance, he/she may be tempted to use the so-called "free lunch" strategy, \textit{i.e.}, take advantage of the low epidemic activity (due to others' efforts) while not contributing to the effort himself/herself.
However, population benefits are at stake because the epidemic dynamics is induced by the aggregation of individual decisions. In practice, an equilibrium is formed between the individuals and we analyze in this work this equilibrium.

From the technical point of view, we study a "Mean Field Game". Introduced by Lasry and Lions in \cite{lasry2006jeux, lasry2006jeux2, lasry2007mean} and independently by Huang, Caines, and Malhamé \cite{huang2006large, huang2007invariance, huang2007large, huang2007nash}, Mean Field Games focus on the derivation of a Nash equilibrium within a population containing an infinite number of individuals, see \cite{carmona2018probabilistic} for a complete mathematical description. Such asymptotic viewpoint simplifies the game theoretic analysis of the interactions among the population, as typically the impact of each individual over the entire population can be neglected. It is worth pointing out that the optimal strategy in a Mean Field game provides approximate Nash equilibrium for similar games involving a finite number of players (see \cite{carmona2018probabilistic}). Fields of applications of Mean Field games, mentioned in \cite{gueant2011mean}, include energy \cite{elie2019mean}, finance \cite{carmona2013mean, elie2020large}, crowd modeling \cite{achdou2016mean}, and also epidemic dynamics  \cite{bauch2004vac,laguzet_individual_2015,laguzet2016equilibrium,hubert2018nash,salvarani_optimal_2018}. These latter works focus on the impact of individual vaccination decisions on the dynamics of the epidemic. In this paper, we follow a similar approach but study the impact of individual decisions concerning distancing and isolation, in an epidemic dynamics where no vaccination is (yet) available.

Optimal control of epidemic is often modeled through a combination of isolation and vaccination strategies. Early work of \cite{abakuks1973optimal,abakuks1974optimal} focuses on the optimal vaccination timing of individuals, while combined with cost free instantaneous isolation strategy. More realistic impacts and constraints on the quarantine and isolation strategies  have been  considered in \cite{morton1974optimal,wickwire1975optimal} or \cite{behncke00,hansen11}. The main objective of such policy is to control the reproduction number of the epidemic (see \cite{perasso2018introduction}) together with limiting the social and economic impact of such policy. 
An other part of the literature (see \cite{fenichel2011adaptive, sahneh2012existence, rizzo2014effect,donofrio_vaccinating_2007,donofrio_fatal_2008,buonomo_global_2008,wang_statistical_2016} for example) tries to model and to take into account the individual behavioral response to isolation policy in such context. Such modeling of individual response is of course greatly influenced by  cultural habits together with societal, economic or religious need for social interactions. The addition of such individual based feedback effect on quarantine governmental policy is key in the modeling of optimal control for epidemic dynamics. The main outcome of this paper is to provide an easily tractable modeling approach for such a purpose. 

We address in this work the question of epidemic control using an individual based modeling approach, and ask the question of how does the cost of a social distancing strategy disseminate among individuals. Each individual chooses optimally his/her contact rate with others, striking a balance between the cost of being infected by the virus, and the cost induced by reducing social interactions.  On one hand, the epidemic dynamics is induced by the aggregation of all individual contact rate decisions. On the other hand, the epidemic level influences the contamination probability of each individual and hereby their own contact rate. We prove that a Mean Field Nash equilibrium is reached among the population and we quantify the impact of individual choice between epidemic risk and other unfavorable outcomes. Numerical experiments show that the  self isolation equilibrium strategy is characterized by a quick contact rate reduction followed by a slower return to the usual social interaction level. It allows to reduce significantly the proportion of infected individuals required in order to achieve herd immunity.

We also compare the induced epidemic dynamics to a situation where a global planner (typically a strongly empowered government) is able to control all the interaction rates among individuals. We observe numerically that the Mean Field Nash equilibrium provides a sub-optimal contact rate strategy in comparison to the societal optimum, and measure the corresponding so-called \textit{cost of anarchy}. We observe in particular that the divergence between the individual and societal strategies  happens  after  the  epidemic  peak  but  while  significant  propagation is  still  underway. By taking into account the individual responses to different cost structures, our modeling approach can provide insights on the impact of political decisions concerning the costs induced by contact rate reduction.

The paper is organized as follows. We first present in Section \ref{Section_model} the SIR model used to describe the COVID-19 2020 epidemic dynamics, together with the control problem faced by each individual. We prove in section \ref{Section_MeanField} the existence of a Mean Field Nash equilibrium in our setting and present a numerical algorithm for its approximation. Section \ref{Section_numerics} focuses on the numerical experiments and presents the computing strategy, and the results in terms of cost of anarchy and sensitivity of the optimal contact rate strategy to the main key modeling parameters. It provides in particular a comparison with a global planner setting. In order to take into account for the relatively long latent phase of the COVID-19 disease, we extend our study in Section \ref{sec:SEIR} to an SEIR epidemic model, and present some numerical results. Section \ref{Section_maths} provides the technical parts of the mathematical analysis associated to our study. Finally, Section \ref{sec:conclusion} summarizes the main outputs of the paper and discusses their limits and potential extensions.

\section{Individual based modeling of the epidemic dynamics}\label{Section_model}

\subsection{The model}

\begin{figure}[h!]
\begin{center}
    \includegraphics[width=0.55\textwidth]{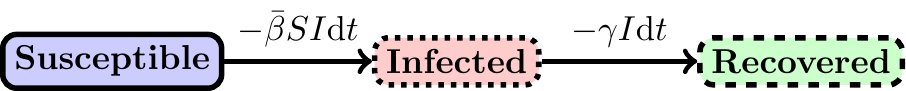}
\caption{Graphical illustration of the SIR model.}
\label{fig:SIR}
\end{center}
\end{figure}

The dynamics of the epidemic is modeled by the standard SIR (Susceptible - Infected - Recovered) compartment model represented in Figure \ref{fig:SIR}. We refer to \cite{Anders2010,capasso_mathematical_1993,capasso_generalization_1978} for additional details on this model, as well as the description of many other mathematical epidemic propagation models (and to \cite{ng2003ado,turinici_sars_2006,danchin_new_2020,volpert_coronavirus_2020} for specific coronavirus models). During the epidemic propagation, each individual can be either "Susceptible", "Infected" or "Recovered", and $(S_t,I_t,R_t)$ denotes the  proportion of each category at time $t \ge 0$. The group of "Recovered" represents the collection of individuals whose behavior does not impact the transmission of the virus anymore. This includes the individuals who potentially did not survive the epidemic, as well as the individuals tested positive and isolated in a perfect quarantine. Besides, we hereby implicitly assume that the immunity acquired by the "Recovered" lasts for ever. 
As a first step,
we choose to restrict our analysis to a simple SIR model rather than considering more complex (and realistic) models, in order to focus our analysis on the role of individual decisions aggregation on the epidemic dynamics. 
Nevertheless, our study can be extend to more complicated epidemic models. In particular, an extension to the SEIR compartment model is presented in Section \ref{sec:SEIR}.

More precisely, the (continuous-time) evolution of the disease is described by the following equations:
\begin{equation}\label{sys:SIR}
\begin{cases}
    \drm S_t = - \bar\beta_{t} S_{t} I_{t}  \drm t\\
    \drm I_{t}  = \bar\beta_{t} S_{t} I_{t} \drm t - \gamma I_{t}  \drm t\\ 
    \drm R_{t} = \gamma I_{t}  \drm t,
\end{cases}
\end{equation}
with a given initial compartmental distribution of individuals at time $0$ denoted $(S_0,I_0,R_0)$, which is supposed to be known.
We assume that initially $S_{0} + I_{0} + R_{0} = 1$, and observe that the system \eqref{sys:SIR} possesses a conservation law, \textit{i.e.}, for all $t\ge 0$, $S_t + I_t + R_t = 1$.\footnote{The SIR model can be extended to include births and deaths. Nevertheless, given that the duration of the COVID-19 pandemic is assumed to, hopefully, be relatively short with respect to the life expectancy at birth in most of the concerned countries, the demographic dynamics is not relevant and will not be taken into account.} As the dynamics of the epidemic is not impacted by the size of the class $R$, we can restrict our focus on the evolution of the proportion $(S,I)$ of both susceptible and infected.

The model described by \eqref{sys:SIR} involves two parameters, $\gamma$ and $\bar \beta$. The constant parameter $\gamma > 0$ is exogenous and identifies  the recovery rate, \textit{i.e.}, it represents the inverse of the lenght (in days) of the contagious period (or time before strict isolation). In our framework, the key parameter is the transmission rate of the disease, denoted $(\bar\beta_t)_{t\ge 0}$, which is considered to be endogenous and time-dependent, contrary to more classical SIR models. To focus our study on this transmission rate, we assume that the parameter $\gamma$ is fixed and known.

The transmission rate $\bar\beta$ depends essentially on two factors: the disease characteristics and the contact rate within the population. We will denote by $\beta_0$ the constant initial transmission rate of the disease, \textit{i.e.}, without any control measures or effort from the population. Although the society cannot modify the disease characteristics, it can produce (possibly strong) incentives to each individual to reduce his/her contact rate with other individuals in the population. This mitigation strategy (lockdown) has been chosen by many  countries during the Covid-19 epidemic propagation, some countries starting strict measures in January (China) while other countries in February, March and so on; this is implemented in order to slow down the epidemic propagation. However, for each individual, reducing the contact rate with others comes at a cost, as described in the introduction (health hazards, psychological pressures, loss of social relationships, income uncertainty...) and it may even expose the individual to yet unknown risks. The latter is especially true for long epidemics that require substantial efforts from the individuals. On the other hand, if everybody else lowers his/her own contact rate then the epidemic diffusion will be vanishing and a given individual, having this knowledge, can act as a free-rider and effectively increase his/her contact rate without too much risk.

Finally, the global epidemics propagation rate $\bar\beta$ of the society follows from the aggregation of all individuals contact rates, so that an equilibrium is reached between all contact rates $\beta$ chosen by each individual and the overall epidemic transmission rate $\bar\beta$ in the society.

\subsection{Individual viewpoint: contact rate optimization}

We focus in this section on the individual perspective on the epidemic control. Consistent with the literature, we suppose that the population can be partitioned into several collections of identical individuals (here the "Susceptible", "Infected and infectious" and "Removed" classes), so that each individual in a given collection takes the same decisions as other individuals in the same class. This implies that we can consider a "representative" individual, which is an arbitrary individual in a given class. As we are interested in the unfold of the COVID-19 pandemic (as opposed to an endemic disease such as measles, mumps, rubella, pertussis, ...) we will consider a finite time horizon $T>0$, large enough so that the epidemic is over at time $T$.\footnote{Considering a finite horizon time also allows some technical simplifications of the mathematical analysis, mainly for the proof of the existence of a Nash equilibrium, without sacrificing qualitative conclusions. Nevertheless, the results can be extended to an infinite horizon, with relevant  technical adaptations. Moreover, since the numerical simulations can only tackle finite times, this choice has the advantage to use an unified framework for both contributions.}

The probability of infection of the representative individual depends on both his/her own contact rate $\beta$ with the population, together with the proportion $I_t$ of infectious individuals in the population. More precisely, let $\tau$ denote the random infection time of the representative individual. His/Her probability $P^\beta_t = \P(\tau \leq t)$ of being infected before time $t$, when using contact rate $\beta$, has the following dynamics (see \cite{laguzet_individual_2015,laguzet2016equilibrium}): 
\begin{align}\label{eq:dynamique_tau}
\drm P^\beta_t = \beta_t I_t \big(1-P^\beta_t \big) \drm t , \; \text{ for } \; t \in [0,T],
\end{align}
where $I_t$ is the proportion of (contagious) infected at time $t$, whose dynamics is driven by the population contact rate $\bar\beta$ (as described in \eqref{sys:SIR}).

We assume that, while in the Susceptible class, \textit{i.e.},  before the (possible) infection time $\tau$, the representative individual can choose his/her contact rate $\beta_t \in [\bmin, \beta_0]$ (for all $t \leq \tau$). Recall that $\beta_0$ represents the transmission rate of the disease without any control  measures, while $\bmin >0$ is the lowest possible contact rate value that can be attained by the representative individual. The efforts of an individual for decreasing social interactions, in order to lower the transmission rate of the virus from $\beta_0$ to some $\beta \in [\bmin, \beta_0]$, induce an instantaneous cost, represented by a decreasing function $c : [\bmin, \beta_0] \to \R^+$\footnote{See Assumption \ref{ass:cost} for the rigorous statement of the hypothesis made on the cost function $c(\cdot)$.}.

Following \cite{laguzet_individual_2015}, we assume that the global cost (seen from time $t=0$) of an individual, while infected at time $\tau$ and choosing a dynamic contact rate $(\beta_t)_t$, is defined by:
\begin{align}\label{eq:cout_tau}
    C (\beta, \tau) := \int_0^{\tau \wedge T} c \big(\beta_s \big) \drm s + r_I \mathds 1_{\tau \leq T},
\end{align}
where $\tau \wedge T$ denotes for the minimum between $\tau$ and $T$.

More precisely, when reducing his/her social interactions on the time interval $[0,T]$, the representative individual faces a global cost given by the sum of:
\begin{enumerate}[label=$(\roman*)$]
    \item a cost for reducing social interactions until being infected, represented by:
    \begin{equation}
    \int_0^{\tau \wedge T} c \big(\beta_t \big) \drm t,
    \end{equation} 
    meaning that, if the individual is not infected during the period $[0,T]$, his/her efforts are costly until $T$. On the contrary, if the individual is infected before $T$, his/her effort are only costly before $\tau$.
    \item a cost incurred at the infection time $\tau$, defined by $r_I \mathds 1_{\tau \leq T}$, where $r_I$ denotes the unitary cost of infection as quantified by the representative individual, together with any other costs that the presence in the "Infected" class implies. Note that $r_I$ is taken here as a constant which does not depend on the pandemic evolution neither on the individual control measures that are specific to the "Infected" class.
\end{enumerate}
We do not specify here the nature of these previous costs, both have health components but may also include other types of costs. We refer the reader to the literature on the QALY/DALY scales for further details~\cite{zeckhauser1976now, anand1997disability, sassi2006calculating}.



Hence, the expected cost of the representative individual, seen from time $t=0$, when using individual contact rate $\beta$ while the population contact rate is $\bar\beta$, is given by the following:
\begin{align}\label{eq:objectif}
    \Ck (\beta, \bar \beta) := \E \big[ C (\beta, \tau) \big] = \E \bigg[ \int_0^{\tau \wedge T} c (\beta_s) \drm s + r_I \mathds 1_{\tau \leq T} \bigg],
\end{align}
where the distribution of $\tau$ is given by \eqref{eq:dynamique_tau} and thus indirectly driven by $\bar \beta$. 

In order to determine his/her optimal contact rate $\beta^*$ while the population rate driving $(I_t)_t$ is $\bar\beta$, the representative individual faces the following minimization problem:
\begin{align}\label{eq:minimization_pb}
    \min_{\beta \in \Bc} \Ck (\beta, \bar \beta),
\end{align}
where the set $\Bc$ of admissible contact rate strategies is defined by:
\begin{align}\label{eq:definitionensembleB}   
\Bc := \{ \beta:[0, T] \to [\bmin,\beta_0], \; \beta \textrm{ measurable} \}. 
\end{align}

\subsection{Societal viewpoint: induced epidemic dynamics}

In the previous subsection, we described the situation from the point of view of a representative individual. We assumed that each individual chooses his/her own contact rate $\beta$ in order to minimize the associated cost $\Ck (\beta, \bar \beta)$, while $\bar \beta$ represents the population's contact rate. In our (Mean Field games) framework, one individual taken alone has no influence on the pandemic dynamics when choosing his/her contact rate. However, the global epidemic propagation rate $\bar\beta$ of the society is induced by the aggregation of all individuals contact rates. Besides, the behavior of the representative individual described in the previous section, is the one used by all individuals assumed to be identical. 

Hence, following the prescriptions of the Mean Field games framework, an equilibrium is sought between the contact rate $\beta$ chosen by each individual and the overall epidemic transmission rate $\bar\beta$ observed at the society level. More precisely, we look for an equilibrium among all individuals in the population, in the sense of the following definition.
\begin{defn}[Mean Field Nash equilibrium]\label{def:Nash_eq}
The contact rate $\beta^\star\in\Bc$ together with the epidemic dynamics $(S^\star,I^\star, R^\star)$ is a Mean Field Nash equilibrium if the following relations are satisfied:
\begin{enumerate}[label=$(\roman*)$]
    \item \textbf{Individual rationality}: it is optimal for the representative individual to choose the contact rate $\beta=\beta^\star$ when the epidemic dynamics is $(S^\star,I^\star, R^\star)$; 
    \item \textbf{Population consistency}: whenever the population contact rate is $\bar\beta=\beta^\star$, the induced epidemic dynamics is given by $(S^\star,I^\star, R^\star)$. 
\end{enumerate}
\end{defn}
Identifying such an equilibrium boils down to a fixed point property of the optimal best response function of the representative individual as described in the following Section. 

\section{Mean Field Nash equilibrium and Numerical approximation}\label{Section_MeanField}

\subsection{Mean Field Nash equilibrium}

As emphasized in the previous section, the contact rate chosen by each individual is impacted by the epidemic dynamics through the proportion of infected $(I_t)_t$, while the number of infected is a direct consequence of the aggregation of all individuals' behavior. We thus look for an equilibrium in such context, \textit{i.e.}, in the sense of Definition \ref{def:Nash_eq}.
In fact, finding a Mean-Field Nash equilibrium is equivalent to identifying a fixed point of the so-called best response function, which provides an optimal individual contact rate $\beta^*\in \Bc$ (or a set of optimal strategies if non-uniqueness of the optimal strategy) in response to a population contact rate $\bar\beta$. 

We therefore introduce  the best response function $\Tc$, which associates to any societal transmission rate $\bar \beta \in \Bc$, the optimal individual contact rate $\beta^\star$:
\begin{align}\label{eq:definitiontau}
    \Tc : \bar \beta \in \Bc \longrightarrow  \beta^\star \in \Big\{ \argmin_{\beta \in \Bc} \Ck (\beta, \bar \beta) \Big\}.
\end{align}
As mentioned earlier, although we expect to find at least one fixed point to this application, in general $\Tc$ is a multi-valued mapping. We make the following assumption:
\begin{ass}\label{ass:cost}
The cost function $c$ is decreasing, two times differentiable with continuous second derivative (\textit{i.e.}, of $\mathcal{C}^2$ class) and the following holds:
\begin{align}\label{eq:hypstrictconvexc}
\inf_{\beta \in [\bmin,\beta_0]} c''(\beta) > 0.
\end{align}
\end{ass}
Most of the results that follow also hold under much less demanding assumptions, but at the price of increasing technicalities.
We first state the theoretical result concerning the existence of a unique best response strategy in $\Tc(\bar \beta)$ for a given $\bar \beta \in \Bc$.
\begin{lem}\label{lemma:uniquebeta}
Suppose $\bar \beta \in \Bc$ is fixed. Under \textnormal{Assumption \ref{ass:cost}}, there exists a unique $\beta^{\star,{\bar \beta}} \in \Bc$ satisfying
\begin{align}
    \Ck (\beta^{\star,{\bar \beta}}, \bar \beta) \leq \Ck (\beta, \bar \beta), \; \text{ for all } \; \beta \in \Bc.
\end{align}
\end{lem}
We are now in position to state the main mathematical result of the paper, concerning the existence of a Mean Field Nash equilibrium. 
\begin{thm}\label{thrm:existence}
Under \textnormal{Assumption \ref{ass:cost}}, the model \eqref{sys:SIR} admits a Mean Field Nash equilibrium $\beta^\star \in \Bc$ together with $(S^{\beta^\star},I^{\beta^\star},R^{\beta^\star})$, in the sense of \textnormal{Definition \ref{def:Nash_eq}}, \textit{i.e.}, for any $\beta \in \Bc$,
\begin{align}
    \Ck (\beta^\star, \beta^\star) \leq \Ck(\beta, \beta^\star).
\end{align}
\end{thm}

Note that the result of Theorem \ref{thrm:existence} only informs on the existence of an equilibrium and not to its possible uniqueness. The uniqueness of a Mean Field equilibrium has been recognized as a difficult problem from the very beginning of the MFG theory (see counter-examples in \cite[remark after Theorem 2.2]{lasry2006jeux}) and very little theoretical guidance exists to support such a claim. Among the  methods that can allow to obtain uniqueness one can list the monotony assumptions (see \cite{lasry2007mean,lasry2006jeux,lasry2006jeux2}) or the evolution dynamics (see \cite{turinici_metric_2017}) but none seems to apply directly here. Moreover, with few exceptions, the epidemic Mean Field equilibrium has rarely been proved to possess systematic uniqueness properties. 
For sake of clarity, the proofs of both Lemma \ref{lemma:uniquebeta} and Theorem \ref{thrm:existence} are postponed to Section \ref{sec:proofs}.

\subsection{Numerical approach}
\label{sec:numerical}

We describe below the methodology used in the numerical experiments in order to approximate the Mean Field Nash equilibrium $\beta^\star$. Recall that the cost for the representative individual when choosing a contact rate $\beta$ while the societal transmission rate is $\bar \beta$, is given by $\mathfrak{C}(\beta, \bar \beta)$.

The metric equilibrium flow approach introduced in  \cite{turinici_metric_2017} (to which we refer the reader for rigorous mathematical transcription of the objects below, see also  \cite{salvarani_optimal_2018} for an application) prescribes the following iterative procedure in order to reach an equilibrium: choose a pseudo-time step $h>0$ and define iteratively  $\beta^{n+1}$ as a minimizer of the following functional
\begin{align}\label{eq:defflow1}
\beta \mapsto \frac{d(\beta,\beta^n)^2}{h} + 
\mathfrak{C}(\beta,\beta^n).
\end{align}
Here, $d(\cdot,\cdot)$ is a geodesic distance on the space $\mathcal{B}$ of all possible individual choices $\beta$. 
Note that when $\mathcal{B}$ is a Hilbert space, 
$\mathfrak{C}(\cdot,\cdot)$ is smooth enough, 
and $d(\cdot,\cdot)$ is the distance induced by the canonical norm, the fact that $\beta^{n+1}$ 
is a minimizer of the functional in \eqref{eq:defflow1} implies:
\begin{align}\label{eq:defflowimplicit}
\beta^{n+1} =\beta^{n} - h \nabla_1  \mathfrak{C}(\beta^{n+1},\beta^n),
\end{align}
where $\nabla_1$ denotes for the derivative with respect to the first argument.

Relation \eqref{eq:defflowimplicit} is similar to the JKO scheme used in one variable gradient flows, see \cite{ambrosio_gradient_2008}. The problem with \eqref{eq:defflowimplicit} is that it is implicit and thus not compatible, in this form, with numerical computation. In practice, when the vectorial structure on $\mathcal{B}$ is compatible with the topological structure, one can propose an iterative procedure to find $\beta^{n+1}$ solution of 
\eqref{eq:defflowimplicit}:
start with 
$\beta^{n+1,0} = \beta^{n}$ and iterate for $\ell \ge 0$
\begin{align}\label{eq:defflowimplicitpicard}
\beta^{n+1,\ell+1} =\beta^{n} - h \nabla_1  \mathfrak{C}(\beta^{n+1,\ell},\beta^n). 
\end{align}
It is standard to see that for $h$ small enough and in presence of (e.g. Lipschitz) regularity of $\nabla_1  \mathfrak{C}$ with respect to its first argument, the iterations are guaranteed to converge, by a Picard fixed point argument, to the (unique) solution of \eqref{eq:defflowimplicit}.
However, for numerical convenience, in practice only $L$ iterations of
\eqref{eq:defflowimplicitpicard} are performed, and for our numerical simulations, $L=1$ worked just fine. Therefore, we implement the following proxy for the minimization in  
\eqref{eq:defflow1}:
\begin{align}\label{eq:defflowexplicit}
\beta^{n+1} = \beta^{n} - h \nabla_1  \mathfrak{C}(\beta^n,\beta^n).
\end{align}
When $\mathcal{B}$ is a Hilbert space and forgetting any possible regularity issues, we obtain an Explicit Euler discretization of the  equilibrium flow 
defined in \cite{turinici_metric_2017}. Thus, we can expect that $\lim_{n\to \infty} \beta^{n}$ will be the equilibrium we want to compute.

\section{Numerical experiments}\label{Section_numerics}

\subsection{Choice of parameters}

The following numerical experiments are done using a contact rate reduction cost function defined by: 
\begin{align}\label{eq:cost_effort}
    c(b) := \dfrac{\beta_0}{b} - 1, \; b \in [\beta_{min}, \beta_0]\,.
\end{align}
Recall that the parameter $\beta_{min}$ represents the minimal achievable contact rate by a representative individual, while $\beta_0$ denotes the usual contact rate used before the beginning of the lockdown measures. In other words, $\beta_0$ represents the transmission rate of the disease without any isolation effort of the population. 
The shape of the cost function encompasses the increasing difficulty to bring the contact rate closer to zero. Conversely, without effort of the individual at time $t$, meaning that $\beta_t = \beta_0$, the associated cost $c(\beta_t)$ is equal to zero. In addition, note that this function $c$ satisfies Assumption \ref{ass:cost}.

The set of parameters used in the experiments are provided in Table \ref{tab:params}. The associated reproduction number $\mathcal R_0$ without isolation measure, commonly defined by $\mathcal R_0 := \beta_0 / \gamma$ in the literature on epidemic models, is equal to  $2.0$ in our framework, and is thus in the confidence interval of available data \cite{li_early_2020}. The parameter $\gamma$ corresponds to the inverse of the virus contagious period see \cite{anderson2009mat}. We assume that the initial proportion of infected $I_0$ in the overall population is $1\%$ at time $0$, when the contact rate optimization starts. We set the cost $r_I$ incurred by an infected individual to $r_I=300$. We recall that  this cost is not necessarily expressed in terms of money, but can also be medical side effects or general morbidity (see \cite{zeckhauser1976now, anand1997disability, sassi2006calculating} for an introduction on QALY/DALY) and is relative to the definition of $c(\cdot)$.

\renewcommand{\arraystretch}{1.3}

\begin{table}[h]
\centering
\begin{tabular}{|c|c|c|c|c|c|c|}
\hline
    $(S_0,I_0,R_0)$  &  $\gamma$ & $\beta_0$  & $\beta_{min}$  & $\underline{\beta}$ & $T$ & $r_I$ \\
    \hline
    $(0.99, 0.01, 0.00)$ & $1/10$ & $0.20$  & $0.05$ & $0.14$ &  $360$ days
    & $300$  \\
    \hline
\end{tabular}
\vspace{0.5em}
\caption{Set of parameters for the numerical experiments}
\label{tab:params}
\end{table}

Finally, recall that there is considerable uncertainty in the medical literature on the choice of all parameters described above, so the sensibility to the values chosen has been tested in Section \ref{sec:sensibility}.

\subsection{Mean Field Nash equilibrium}

The numerical approximation of the Mean Field Nash equilibrium is obtained using the algorithm described in Subsection \ref{sec:numerical}. The initial guess $\beta^0$ is taken constant $\beta^0(t)=\beta_0$, $\forall t \ge 0$.\footnote{We tested other initial guesses and all the convergence rate was the same. Moreover, in all cases, the same equilibrium is obtained.} 

\begin{figure}[!ht]
\includegraphics[scale=0.5]{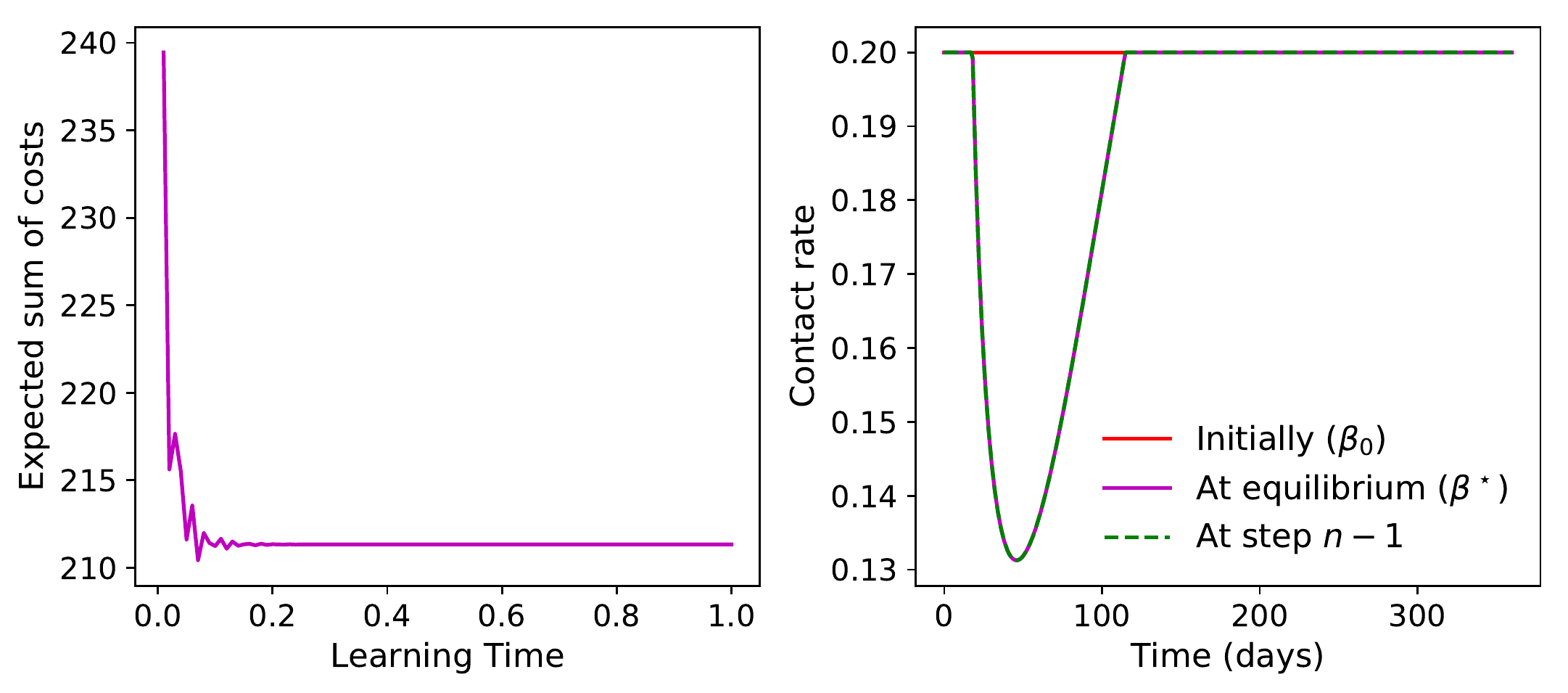}
\centering
\caption{Convergence of the algorithm to the Mean Field Nash equilibrium of the model \eqref{sys:SIR}, using the parameters described in Table \ref{tab:params}. Left: global cost $\Ck$ in terms of the learning time. Right: Mean Field Nash equilibrium contact rate $\beta^\star$ (magenta line), identical to the contact rate at the penultimate step of the algorithm (green dashed line) and compared to $\beta_0$ (red line).} 
\label{fig:convergence}
\end{figure}

\begin{figure}[!ht]
\includegraphics[scale=0.5]{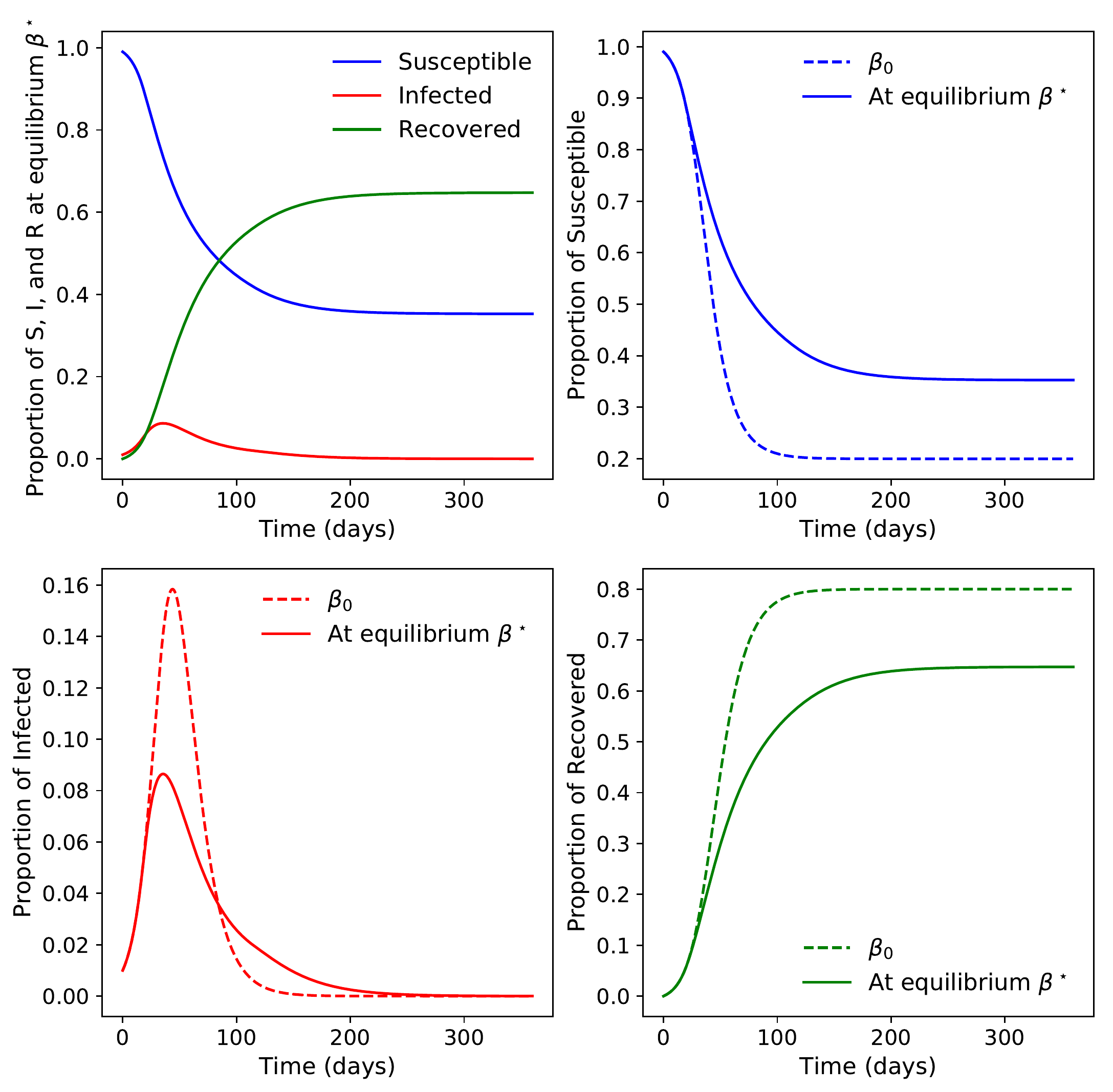}
\centering
\caption{Evolution (in proportion) of Susceptible, Infected and Recover classes, using the parameters described in Table \ref{tab:params}. Solid lines represent the evolution at the Mean Field Nash equilibrium, \textit{i.e.}, with transmission rate $\beta^\star$. The evolution at equilibrium is compared to the epidemic dynamics with constant transmission rate $\beta_0$ (dashed lines).} 
\label{fig:SIR_equilibrium}
\end{figure}

We plot in Figure \ref{fig:convergence} the convergence towards the minimal cost; the decay of the cost is very fast in terms of the learning time $h$. 
The contact rate at equilibrium is represented in the right plot, while the empirical convergence of the algorithm is confirmed by its superposition with the approximate contact rate computed at the previous step. 
At this numerical Mean Field Nash equilibrium, the cost function $\Ck(\beta^\star, \beta^\star)$ is  $211.33$ and provides for each individual a relative gain of $12 \%$ in comparison to the zero effort strategy $\beta_0$. In addition, only $85\%$ of this cost is explained in terms of the wealth impact (instead of $100\%$ for the strategy $\beta_0$), with the remaining $15\%$ being related to the cost of social effort; the probability of being infected over the interval $[0,T]$ decreases 
from $80\%$ in the $0$-effort benchmark scenario,
to $62\%$, see Figure \ref{fig:SIR_equilibrium}. Note that in this case, the minimum attainable infection probability is $50\%$ (as calculated from \cite[Lemma A.1]{laguzet2015global} or from equation \eqref{sys:SIR}).

The equilibrium contact rate $\beta^\star$ is characterized by three major phases, in response to the Mean Field Nash equilibrium epidemic dynamics presented in Figure \ref{fig:SIR_equilibrium}. First, at the beginning of the epidemic, the number $I$ of infected people is relatively low. Individuals make therefore no effort in order to reduce their social interactions and the virus is transmitted at the normal rate $\beta_0$. This leads to a large augmentation in the proportion of Infected, implying a significant increase of the individual's probability of contracting the virus. In response to this, individuals begin to reduce significantly their social interactions, implying a strong reduction of the transmission rate of the disease. Finally, after the epidemic peak, all individuals slowly reduce their effort until the number of infected people is relatively close to $0$.

Figure \ref{fig:SIR_equilibrium} provides a comparison between the SIR dynamics of the Mean Field Nash equilibrium (solid lines), and the one generated by the no-effort strategy $\beta_0$ (dashed lines). Even if driven by self-interest, individual efforts do reduce social interactions population-wise. This can be explained as follows. 
First, we observe that the proportion $R_T$ of recovered at terminal date drops from $80\%$ to $60\%$ when the population is applying the Mean Field Nash equilibrium strategy. This means that the proportion of the population spared by the virus goes from $20\%$ in the case without effort to $40\%$ at equilibrium. Second, one can observe that the infection peak occurring around $t=50$ days is twice less critical at equilibrium, but, as a counterpart, the epidemic lasts longer as the number of infected decreases more slowly after the epidemic peak. As the infection peak is less critical, it limits, and may even prevent, the saturation of the healthcare facilities. Although not represented in our model, this necessarily implies a decrease in the mortality rate of the virus.

\subsection{Impulse control equilibrium}

We now focus on the situation where the set of admissible strategies is restricted to a subset of $\Bc$, denoted by $\bar \Bc$, of particular piece-wise constant strategies:
\begin{align}
    \bar \Bc := \big\{\beta \in \Bc, \;
    \beta(t)=   \underline{\beta} \cdot \mathds 1_{[t_1, t_2]}(t) +
    {\beta_0} \cdot \mathds 1_{[0,T]\setminus [t_1, t_2]}(t);\;  0 \leq t_1 \leq t_2 \leq T \big\},
\end{align}
where we took $\underline{\beta} = 0.14$ as described in Table \ref{tab:params}. This framework encompasses the realistic situation where the instantaneous contact rate of each individual can not be, in practice, chosen within the whole set $\Bc$ and has to  be restricted to a unique control period. The representative individual will optimally select the times $t_1$ and $t_2$, representing respectively the beginning and the end of his/her lockdown period. 

\begin{figure}[!h]
\includegraphics[scale=0.45]{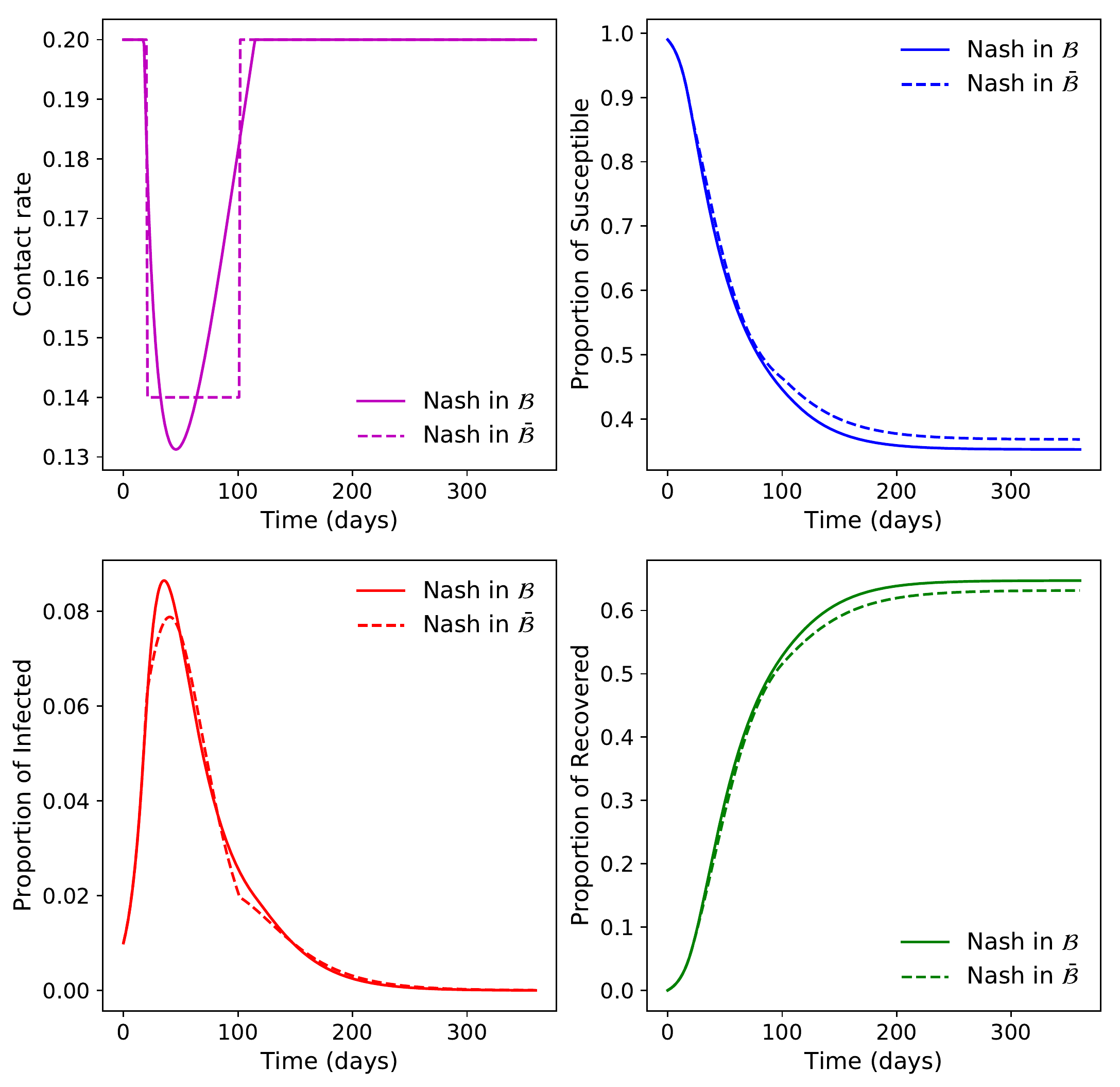}
\centering
\caption{Comparison between Mean Field Nash equilibria in $\Bc$ (plain lines) and  $\bar \Bc$ (dashed line), using the parameters described in Table \ref{tab:params}.} 
\label{fig:SIR_compare_Nash}
\end{figure}

Figure \ref{fig:SIR_compare_Nash} compared the Mean Field Nash equilibria obtained over both sets $\bar\Bc$ and $\Bc$. The equilibrium strategy over $\bar\Bc$ starts isolation measures at time $t_1 = 20$ by decreasing the contact rate from $\beta_0=0.20$ to $\underline{\beta}=0.14$. The duration of the lockdown is $81$ days, after which the contact rate is immediately returning to normal, \textit{i.e.}, $\beta_0$. The cost induced for each individual is around $210.78$, which is lower than the one around $211.33$ associated to the equilibrium over $\Bc$. Moreover, we observe that the induced SIR dynamics provides a lower epidemic size $R_T$  together with a lower proportion of infected at the epidemic peak, hereby reducing the potential mortality rate induced by the virus. This observation enlightens as well how the Mean Field Nash equilibrium $\beta^\star$ in $\Bc$ is not optimal for the society as a whole, as it can be improved by restricting the set of admissible strategies to $\bar\Bc$. This observation leads us to look towards the optimal societal contact rate for the population as a whole.

\subsection{Cost of anarchy}

In our previous equilibrium analysis, each individual is considered to be too small in order to impact the epidemic dynamics of the society and can hence acts in a selfish manner: each individual minimizes his own cost $\Ck( \cdot , \bar \beta)$ in response to the transmission rate $\bar\beta$ of the epidemic. On the other hand, a global planner, \textit{e.g.}, a government with full empowerment, will optimize the global cost of the entire society with respect to the choice of the transmission rate in the society. Namely, the global planner will solve:
\begin{align}\label{eq:minimization_planner_pb}
    \min_{\bar\beta \in \Bc} \Ck (\bar\beta, \bar \beta),
\end{align}
This leads to a  different optimization problem, which is well documented in the literature, see for example \cite{sethi78, behncke00, hansen11, djidjou2020optimal}, and even more largely on the topic of vaccination (see, \textit{e.g.}, \cite{ abakuks1974optimal, morton1974optimal, piunovskiy2008ane, laguzet2015global}).

\begin{figure}[!ht]
\includegraphics[scale=0.5]{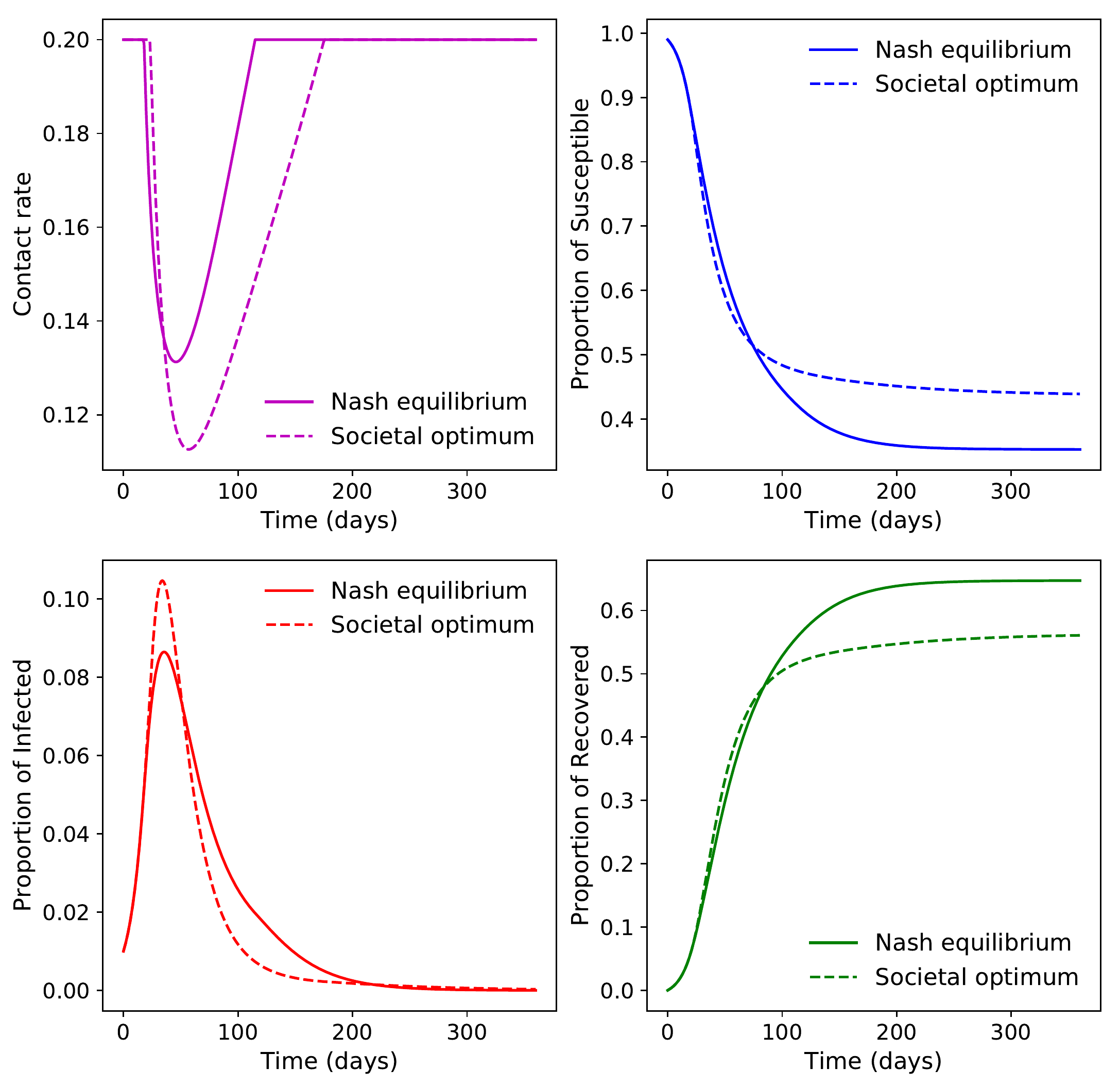}
\centering
\caption{Comparison between two strategies: the Mean Field Nash equilibrium contact rate $\beta^\star \in \Bc$ (solid lines) and the optimal transmission rate for the society (dashed line), using the parameters described in Table \ref{tab:params}.} 
\label{fig:Comparison_optimum_social}
\end{figure}

In our framework, we can compute, through classical optimization procedures (e.g., the Pontryagin principle), the optimal control of the transmission rate from the society point of view. We refer to the literature previously cited for the details on the techniques allowing to find this optimal control 
and we only present here the numerical results. Note that in such context there are several additional, classical, procedures available to compute the optimal control
(see for instance the forward-backward sweep method in \cite[Chapter 4]{lenhart2007optimal}). Nevertheless, in our framework, a slight modification of the previously implemented gradient descent detailed in Subsection \ref{sec:numerical} works just fine.

First, observe that the societal optimal transmission rate imposes 
 larger effort at the beginning of the control period, and relieves these constraints much more slowly. The total control duration period for the societal optimum is $151$ days in comparison to $96$ days for the Mean Field Nash equilibrium, 
 although societal control begins later than in the case of the Nash. 
  Secondly, observe that the optimal transmission rate chosen by the global planner accentuates the already encouraging results obtained with the Mean Field Nash equilibrium on the epidemic dynamics: the epidemic size $R_T$ represents only $55\%$ of the total population.

The societal optimum allows to reach an individual cost around $200.25$, while the Mean Field Nash equilibrium provides a cost valued around  $211.33$. This phenomenon allows to mathematically quantify the so-called "cost of anarchy", induced when letting each individual decide on his/her own, instead of letting a global planner take decisions for the population as a whole. Of course, the societal optimal strategy is not a Mean Field Nash equilibrium: given this optimal transition rate for the society, each individual is tempted to make less effort in reducing his/her own contact rate which will drive away the global rate towards the Mean Field equilibrium.

\begin{figure}[!ht]
\includegraphics[scale=0.5]{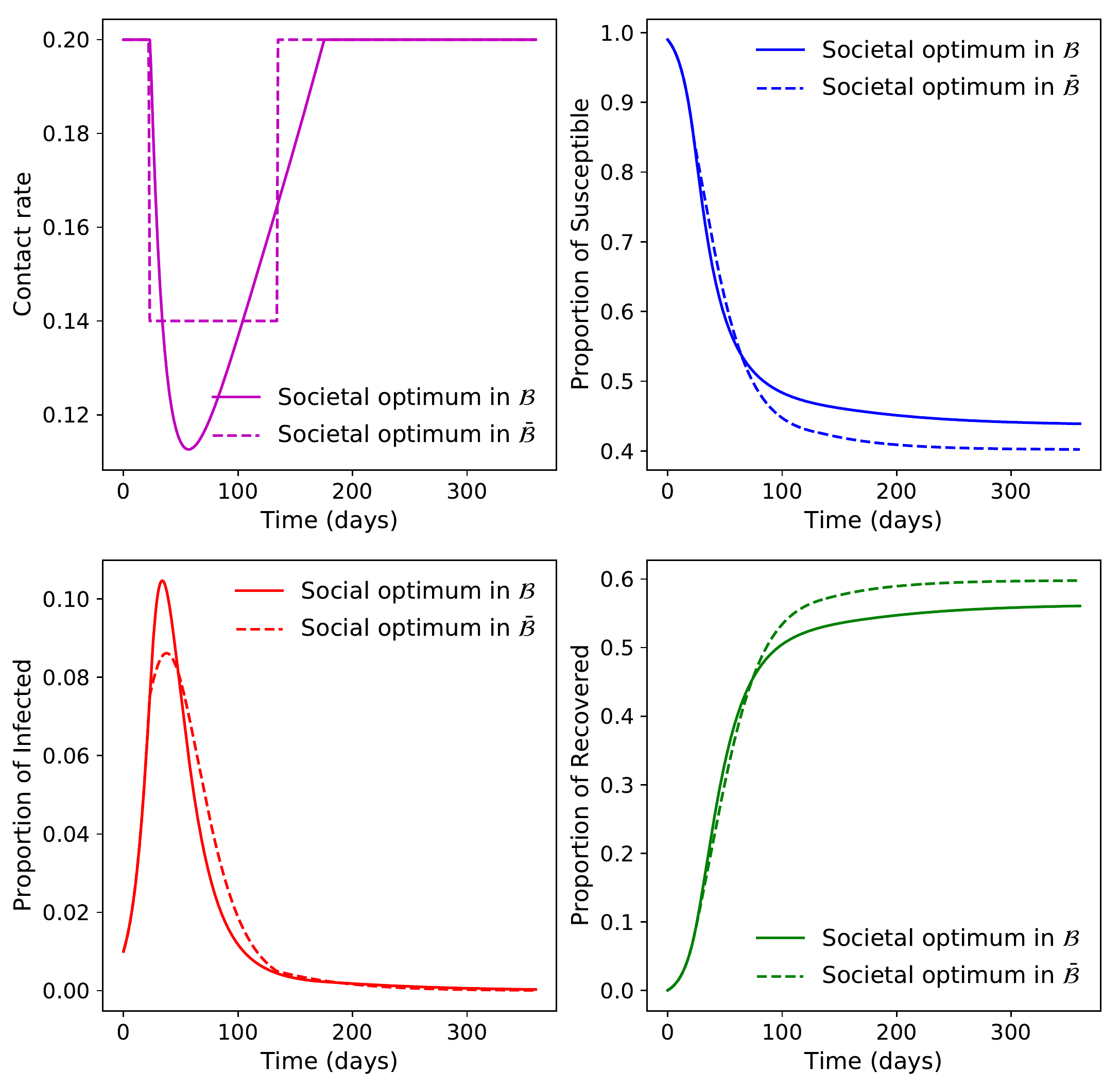}
\centering
\caption{Comparison between two strategies: the societal optimum contact rate in $\Bc$ (solid lines) and the societal optimum contact rate in $\bar \Bc$, using the parameters described in Table \ref{tab:params}.} 
\label{fig:Comparison_optimum_social_bb}
\end{figure}

We also compare in Figure \ref{fig:Comparison_optimum_social_bb} the societal optimal transmission rate in $\Bc$ to the optimal one in $\bar \Bc$, where only one strong effort period is allowed. The  societal-wide optimum control  period in $\bar \Bc$ starts at time $t = 23$ and lasts $111$ days. This induced cost is around $2\%$ higher than the one induced by the optimal social diffusion rate. 
This may allow to quantify the importance of going through a progressive lock-out strategy.

\begin{figure}[!ht]
\includegraphics[scale=0.5]{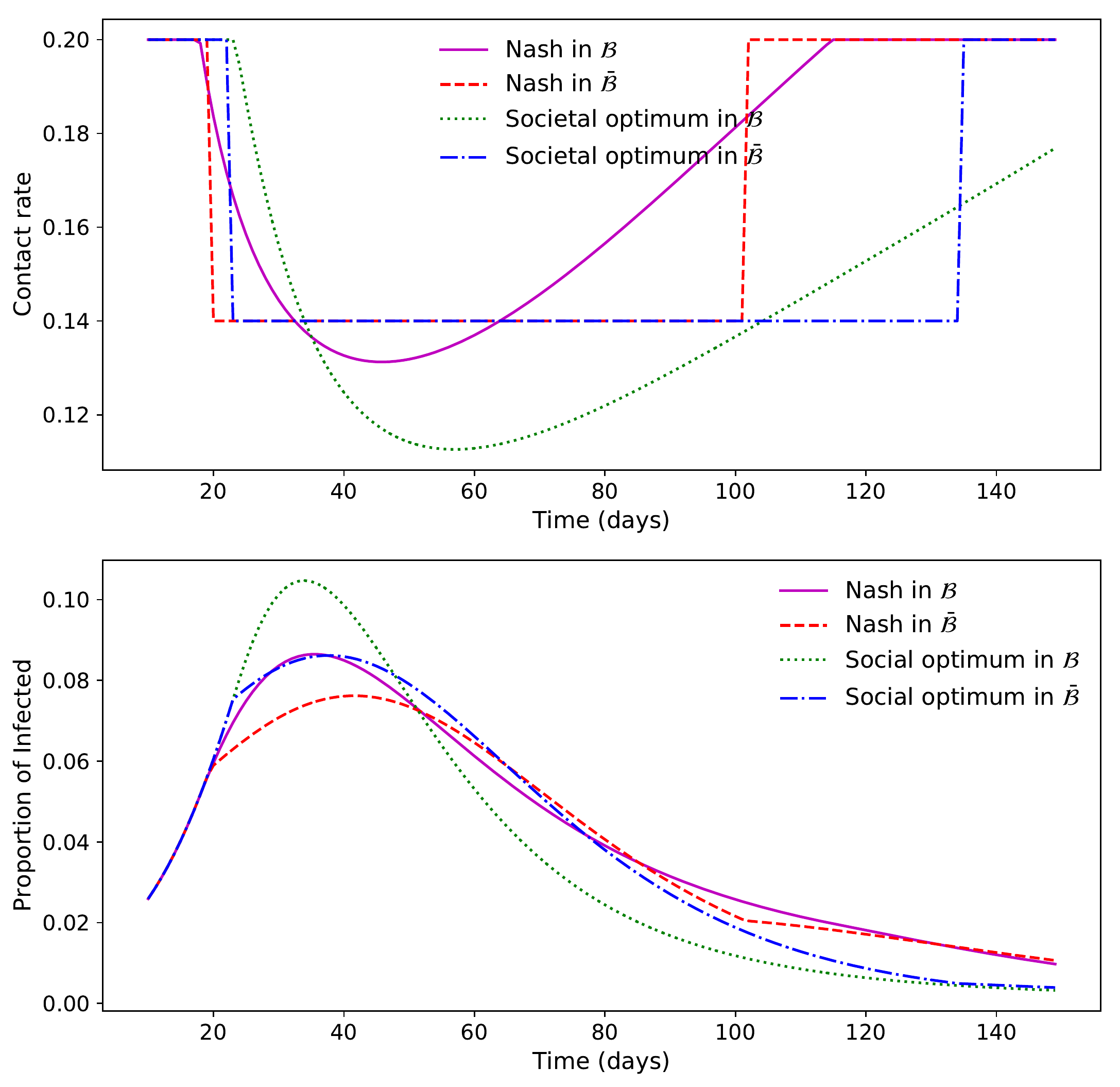}
\centering
\caption{Comparison between the transmission rate and the dynamic proportion of infected on the time interval $[10, 150]$, for the four aforementioned strategies: the Mean Field Nash equilibrium in $\Bc$ (solid magenta lines) and $\bar \Bc$ (dashed red lines), the societal optimum in $\Bc$ (dotted green lines) and $\bar \Bc$ (dash-dot blue lines), using the parameters described in Table \ref{tab:params}.} 
\label{fig:Comparison_all}
\end{figure}

Finally, Figure \ref{fig:Comparison_all} provides a closer look on the four aforementioned strategies, focusing on the time interval $[10,150]$, in order to highlight the control (lockdown) and  control-less (lockout) properties. One can observe that for both Mean Field Nash equilibrium strategies, the control period starts and ends earlier in comparison to the societal optimum situation. Indeed, on one hand, individuals engage in preventive measures by decreasing their interactions earlier than a global planner would recommend, due to fear of the infection spreading. On the other hand, they release their efforts just after the peak of infection, whereas a global planner would recommend maintaining a relatively high level of effort in order to avoid further spreading of the virus. Therefore, while the Nash equilibrium of individuals allows, through premature efforts, to decrease the peak of infection, the socially optimal strategy allows a more rapid decrease in the proportion of infected after the peak, by maintaining intense efforts. Nevertheless, we shall remind that the  global planner should also take into account the possible saturation of the healthcare system, leading to a lower epidemic peak induced by the optimal societal transmission rate.

\subsection{Sensitivity to parameters}\label{sec:sensibility}

The parameters used in the previous numerical experiments are provided in Table~\ref{tab:params} above. Nevertheless,  as the current medical literature still shows considerable uncertainty concerning the numerical values for those parameters, we tested the sensitivity of our findings with respect to the choice of the main parameters of the model. The corresponding figures are provided in Appendix \ref{app:fig_sensibility} for better readability.

\begin{enumerate}[label=$(\roman*)$]
    \item Figure \ref{fig:SIR_sensi_rI} provides the  Nash equilibrium and the epidemic dynamics for three different values for $r_I$, describing the relative effects of the two parts of the costs. As expected, the more costly is the infection for an individual, the more efforts he/she will do in order to limit his social interactions, in order to decrease his/her probability of infection. In particular, reducing the sanitary cost $r_I$ from $350$ to $250$ implies a $20\%$ decrease on the level of the epidemic peak.
    \item Figure \ref{fig:SIR_sensi_beta_0} provides a similar study for three different values of the reproducing number $\mathcal R_0$. For higher $\mathcal R_0$, individuals make more effort in order to reduce their social interactions. Bringing $\mathcal R_0$ from $2.5$ to $1.8$ reduces the epidemic peak by $40\%$ together with reducing the size $R_T$ of the epidemic from $70\%$ to around $60\%$. This result is perfectly understandable since the higher $\mathcal R_0$ is, the higher the probability of being infected without effort is. Each individual limits his social interaction in order to decrease the probability of being infected, and thus diminishes the wealth impact of the epidemic. 
    \item Finally, Figure \ref{fig:SIR_sensi_i0} studies the sensibility of our findings with respect to the initial proportion $I_0$ of infected at time $0$. A higher $I_0$ induces an earlier beginning of the control period, together with a stronger efficacy of it. At terminal date, the total proportion of susceptible remains similar. This implies that the long term effects of a late detection of the epidemic can be compensated by a stronger isolation equilibrium policy. Once again, we omit here the possible negative outcomes induced by the saturation of the health care system. 
\end{enumerate}


\section{The SEIR model and application to COVID-19}\label{sec:SEIR}

It should be noted that the COVID-19 disease is characterized by a relatively long latency phase (as well as many other complex dynamics). To account for this latent phase, where infected individuals are not yet contagious, we can extend our reasoning to an SEIR model, where the class $E$ represents the individuals infected but not yet infectious. The dynamics of the SEIR model are described by the following system:
\begin{equation}\label{sys:SEIR}
\begin{cases}
\drm S_t = - \bar\beta_{t} S_{t} I_{t}  \drm t\\
\drm E_t = \bar\beta_{t} S_{t} I_{t}  \drm t - \alpha E_t \drm t\\
\drm I_{t}  = \alpha E_t \drm t  - \gamma I_{t}  \drm t\\
\drm R_{t} = \gamma I_{t}  \drm t,
\end{cases}
\end{equation}
where $\bar \beta$ still represents the societal transmission rate, and $\alpha > 0$ is a parameter specific to the SEIR model, representing the rate at which an exposed person becomes infectious. The average incubation period is therefore given by $1/\alpha$.

The computations provided in Section \ref{Section_maths} still hold for the SEIR model. Therefore, the application of the numerical scheme described in Section \ref{sec:numerical} is straightforward, and only the numerical results are presented here. The parameters used for the numerical simulations are those provided in \cite{bacaer_modemathematique_2020, dolbeault2020heterogeneous}, and are described in the following table. Other parameters given in Table \ref{tab:params} remain unchanged.

\begin{table}[ht]
\centering
\begin{tabular}{|c|c|c|c|}
\hline
   $(S_0,E_0, I_0,R_0)$  & $\gamma$ & $\alpha$ & $\beta_0$ \\
   \hline
   $(0.99984, 8.81 \times 10^{-5}, 1.88 \times 10^{-5}, 5.35 \times 10^{-5})$ & $1$ & $0.25$ & $2.33$ \\
   \hline
\end{tabular}
\vspace{0.5em}
\caption{Set of parameters for the SEIR model}
\label{tab:params_SEIR}
\end{table}

\begin{figure}[!ht]
\includegraphics[scale=0.5]{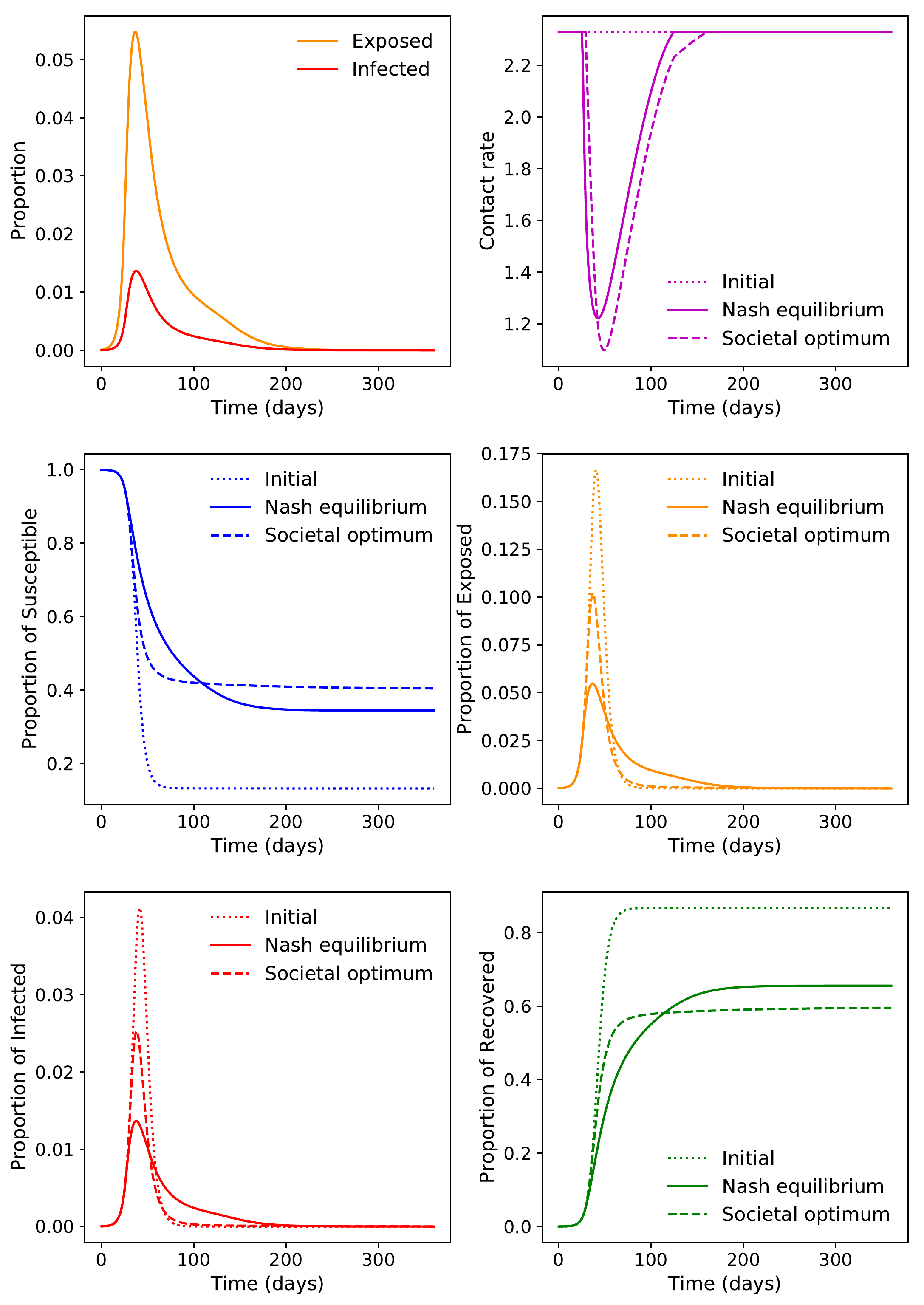}
\centering
\caption{Transmission rate and induced evolution of Susceptible, Exposed, Infected and Recover classes, using the parameters described in Table \ref{tab:params_SEIR}. Solid lines represent the transmission rate and the epidemic evolution at the Mean Field Nash equilibrium, while dashed lines model the societal equilibrium. These two equilibria are compared to the epidemic dynamics with constant transmission rate $\beta_0$ (dotted lines).} 
\label{fig:SEIR_equilibrium}
\end{figure}

The numerical results are provided in Figure \ref{fig:SEIR_equilibrium} and have the same features as for the SIR model. More precisely, the proportion of the population spared by the virus goes from less than $20\%$ in the case without effort to $40\%$ at equilibrium. Moreover, the infection peak occurring around $t=50$ days is three times less critical at equilibrium. This result would limit, and may even prevent, the saturation of the healthcare facilities, and thus implies a decrease in the mortality rate of the virus. However, as a counterpart, the epidemic lasts longer as the proportion of exposed and infected individuals decreases more slowly after the epidemic peak. Finally, and similarly as for the SIR model, the cost of anarchy exists. In particular, the societal optimal transmission rate requires a greater and longer-lasting effort, even if it starts a little later. The optimal transmission rate at the societal level thus improves the rate of recovery from 60\% to 55\%. However, since the effort begins later, the infection peak is higher than for the Nash equilibrium.

\section{Mathematical details}\label{Section_maths}

In this section, we detail the computations and proofs when the epidemic is modeled by a SIR. Nevertheless, these computations still hold if we consider instead an SEIR model, as mentioned in Section \ref{sec:SEIR}. More precisely, since the individual is considered infected as soon as he/she enters class $E$, the probability of being infected in the considered time interval $[0,T]$ remains unchanged. In particular, the cost satisfies the same formula \eqref{eq:formulaCT_2} as for the SIR model, as well as the gradient formula \eqref{eq:gradient_finite}.

\subsection{Probability of infection}

We introduce $(S^{\bar \beta}_{t},I^{\bar \beta}_{t},R^{\bar \beta}_{t})$ in order to denote, at time $t \ge 0$, the solution of the system~\eqref{sys:SIR} with contact rate $\bar \beta$. For all $t \in [0, +\infty)$, we denote by $\varphi_t^{\bar \beta} (\beta)$ the probability that infection occurs before time $t$ for an individual choosing his/her own contact rate $\beta$, while the epidemic evolves according the population's transmission rate $\bar \beta$. Note that $\varphi^{\bar \beta} (\beta)$ is redundant with the notation $P^\beta$ but emphasizes the dependence in $\bar \beta$ of the distribution function of the random infection time $\tau$. 

\begin{lem}
The probability of being infected before time $t \ge 0$, for an individual choosing a contact rate $\beta \in \Bc$, and when the proportion of infected is $I^{\bar \beta}$, is equal to:
\begin{align}\label{eq:proba_infection}
    \varphi_t^{\bar \beta} (\beta) = 1 - \erm^{- \int_0^t \beta_s I_s^{\bar \beta} \drm s }.
\end{align}
\end{lem}
To compute this probability, we  follow  \cite{laguzet_individual_2015,laguzet2016equilibrium}. The Markov chain of an individual, who chooses a contact rate $\beta \in \Bc$ with infected individuals, and whose state at time $t \in [0,T]$ is denoted by $M_t$, is described in terms of the following passage probabilities:
\begin{align}\label{eq:markov_chain}
    \P \big( M_{t + \Delta t} &= S \big| M_{t} = S \big) = 1 - \beta_t I_t^{\bar \beta} \Delta t + o(\Delta t), \nonumber \\
    \P \big( M_{t + \Delta t} &= I \big| M_{t} = S \big) = \beta_t I_t^{\bar \beta} \Delta t + o(\Delta t), \nonumber \\
    \P \big( M_{t + \Delta t} &= I \big| M_{t} = I \big) = 1 - \gamma I^{\bar \beta}_t \Delta t + o(\Delta t), \\
    \P \big( M_{t + \Delta t} &= R \big| M_{t} = I \big) = \gamma I^{\bar \beta}_t \Delta t + o(\Delta t), \nonumber \\
    \P \big( M_{t + \Delta t} &= R \big| M_{t} = R \big) = 1 + o(\Delta t). \nonumber
\end{align}
 The probability of being infected before time $t + \Delta t$ can be written as follows:
\begin{align}\label{eq:proba_infection_def}
    \varphi_{t + \Delta t}^{\bar \beta} (\beta) = &\ \P \big( \exists \tau \in [0, t + \Delta t], \; M_\tau = I \big| M_0 = S \big) \nonumber \\
    = &\ \P \big( M_{t + \Delta t} \in\{I \cup R\} \big| M_{0} = S \big)\nonumber \\
    = &\ \P \big( M_{t + \Delta t} \in\{I \cup R\}  \big| M_t = S \big) \times \P \big( M_{t} = S \big| M_{0} = S \big)  \\
    &+ \P \big( M_{t + \Delta t} \in\{ I \cup R\} \big| M_t \in\{I \cup R\} \big) \times \P \big( M_{t} \in\{I \cup R\} \big| M_{0} = S \big). \nonumber 
\end{align}
Noting that
\begin{align}
    &\ \P \big( M_{t + \Delta t} \in\{ I \cup R\} \big| M_t = S \big) = \P \big( M_{t + \Delta t} = I \big| M_t = S \big) =  \beta_t I_t^{\bar \beta} \Delta t + o(\Delta t), \nonumber \\
    &\ \P \big( M_{t + \Delta t} \in\{ I \cup R\} \big| M_t \in\{ I \cup R\} \big) = 1 + o(\Delta t),
\end{align}
and replacing in \eqref{eq:proba_infection_def}, we obtain:
\begin{align}\label{eq:varphi_2}
    \varphi_{t + \Delta t}^{\bar \beta} (\beta)
    = &\ \beta_t I_t^{\bar \beta} \Delta t \times \big(1-\varphi_{t}^{\bar \beta} (\beta) \big) + 1 \times \varphi_{t}^{\bar \beta} (\beta) + o(\Delta t).
\end{align}
Using the fact that $\varphi_{0}^{\bar \beta} (\beta) = 0$, and letting $\Delta t \rightarrow 0$, the previous equation naturally implies the formula \eqref{eq:proba_infection}.

\subsection{Computation of the cost}

Recall that given a finite time horizon $T$, the expected cost of an individual is defined by \eqref{eq:objectif}.
In other words, we have:
\begin{align}
    \Ck (\beta, \bar \beta) = \E \bigg[ \mathds 1_{\tau \leq T} \int_0^{\tau} c (\beta_s) \drm s 
    + \mathds 1_{\tau > T} \int_0^{T} c (\beta_s) \drm s
    + r_I \mathds 1_{\tau \leq T} \bigg].
\end{align}
Since the cumulative distribution function of the random variable $\tau$ at time $t$ corresponds to the individual's probability of being infected before time $t$, which is denoted by $\varphi_t^{\bar \beta} (\beta)$, we obtain:
\begin{align}\label{eq:formulaCT}
    \Ck (\beta, \bar \beta) &= 
    \int_0^T \int_0^{t} c (\beta_s) \drm s \drm \varphi_t^{\bar \beta} (\beta)
    + \big( 1 - \varphi_T^{\bar \beta} (\beta) \big) \int_0^{T} c (\beta_s) \drm s
    + r_I \varphi_T^{\bar \beta} (\beta) \nonumber \\
    &= \int_0^T c (\beta_s) \big(1 - \varphi_s^{\bar \beta} (\beta) \big) \drm s
    + r_I \varphi_T^{\bar \beta} (\beta).
\end{align}
Moreover, using Equation \eqref{eq:proba_infection}, we can also write:
\begin{align}\label{eq:formulaCT_2}
    \Ck (\beta, \bar \beta)
    &= \int_0^T \big( c (\beta_t) + r_I \beta_t I_t^{\bar \beta} \big) \erm^{- \int_0^t \beta_s I_s^{\bar \beta} \drm s } \drm t.
\end{align}

\subsection{Gradient of the cost}

In order to obtain the gradient of this cost, we have to compute the Gateau derivative $D_h \Ck$ of $\Ck$ with respect to the first variable $\beta$ in the direction $h$. Using Equation \eqref{eq:formulaCT}, we obtain:
\begin{align}\label{eq:gateau_der_1}
    D_h \Ck (\beta, \bar \beta) := &\ \lim_{\varepsilon \rightarrow 0} \dfrac{1}{\varepsilon} \big( \Ck (\beta + \varepsilon h, \bar \beta) - \Ck (\beta, \bar \beta) \big) \nonumber \\
    = &\ r_I D_h \varphi_T^{\bar \beta} (\beta)
    + \int_0^T \Big( \big(1 - \varphi_s^{\bar \beta} (\beta)  \big) D_h c (\beta_s) - c (\beta_s) D_h \varphi_s^\beta (\beta) \Big) \drm s.
\end{align}
Some preliminary computations of the Gateaux derivatives allow to write:
\begin{align*}
    D_h \varphi_s^{\bar \beta} (\beta) := &\ \lim_{\varepsilon \rightarrow 0} \dfrac{1}{\varepsilon} \Big(
    \varphi_s^{\bar \beta} (\beta + \varepsilon h) - \varphi_s^{\bar \beta} (\beta)
    \Big)
    = \big( 1-\varphi_s^{\bar \beta} (\beta) \big) \int_0^s h_u I_u^{\bar \beta} \drm u, \\
    \text{ and } D_h c (\beta_s) := &\ \lim_{\varepsilon \rightarrow 0} \dfrac{1}{\varepsilon} \Big( 
    c \big( \beta_s  + \varepsilon h_s \big)
    - c (\beta_s) \Big)
    = c' (\beta_s) h_s.
\end{align*}
Replacing in \eqref{eq:gateau_der_1}, we therefore obtain the following relation:
\begin{align}\label{eq:gateau_finite}
    D_h \Ck (\beta, \bar \beta) 
    = \langle h,  \nabla_1  \Ck (\beta, \bar \beta) \rangle_T,
\end{align}
where 
\begin{align}\label{eq:gradient_finite}
    \nabla_1  \Ck (\beta, \bar \beta)
    = I_\cdot^{\bar \beta} \Big( r_I \big( 1-\varphi_T^{\bar \beta} (\beta) \big)
    - L_\cdot(\beta, \bar \beta) \Big)
    + \big(1 - \varphi_\cdot^{\bar \beta} (\beta)  \big) c' (\beta_\cdot),
\end{align}
and 
\begin{align}\label{eq:intermediairy_discount}
    L_s(\beta, \bar \beta) := \int_s^T c (\beta_u) \big( 1-\varphi_u^{\bar \beta} (\beta) \big)  \drm u.
\end{align}
Equation \eqref{eq:gradient_finite} is used for numerical simulation, in particular in the equilibrium flow descent described in Subsection \ref{sec:numerical}. 

\subsection{Proof of the existence of an equilibrium}
\label{sec:proofs}

First we recall that Equation \eqref{sys:SIR}
has a unique solution for any ${\bar \beta}\in \Bc$ (see for instance  \cite{bressanimpulse,rampazzo91,silvacontrolmeasure97}). In particular one can prove that if ${\bar \beta}_n$ is a sequence of functions in 
$\Bc$ converging in $L^1$ (thus also in $L^2$) to some ${\bar \beta}_{\infty}$ then the corresponding solution
$(S_t^{{\bar \beta}_n},I_t^{{\bar \beta}_n},R_t^{{\bar \beta}_n})$ 
of \eqref{sys:SIR}
converges (for any given $t$) to 
$(S_t^{{\bar \beta}_{\infty}},I_t^{{\bar \beta}_{\infty}}, R_t^{{\bar \beta}_{\infty}})$.
Moreover for any 
$\beta \in \Bc$
the solution 
$(S_t^{\bar \beta},I_t^{\bar \beta}, R_t^{\bar \beta})$
is a Lipschitz function of time with Lipschitz constant $L_S \le \beta_0 + \gamma$.

 \medskip

We start with the proof of Lemma \ref{lemma:uniquebeta}. 

\begin{proof}
To this end, we consider that  ${\bar \beta}$ is fixed. 
Define the value function
at time $t$ (compare with formula \eqref{eq:formulaCT_2}):
\begin{align}
\Pi_t &= \inf_{\beta \in \Bc}
 \int_t^T c (\beta_s) \big(1 - \varphi_s^{\bar \beta} (\beta) \big) \drm s
    + r_I \left( [1- \varphi_T^{\bar \beta} (\beta)]/[1 - \varphi_t^{\bar \beta} (\beta)] \right)
    \nonumber \\
& =     
\inf_{\beta \in \Bc}
 \int_t^T  \left( c (\beta_s) + r_I 
 \beta_s I_s^{\bar \beta} \right)
 e^{- \int_t^s \beta_u  I_u^{\bar \beta} \drm u } \drm s.
\label{eq:formulaPit}
\end{align}
This corresponds to the optimal cost of an individual starting at time $t$ in the susceptible class. By invoking standard arguments (see \cite{bardi1997}) one can show that $\Pi_t$ is the (unique) solution of the following Hamilton-Jacobi-Bellman equation:
\begin{align}
    \frac{\drm \Pi_t}{\drm t} = H(t,\Pi_t), \ \Pi_T=0,
\end{align}
where 
\begin{align}
H(t,x) = \min_{y \in [\bmin,\beta_0]} \Big\{
 c (y) + (r_I-x) 
 y I_t^{\bar \beta} \Big\}.
\label{eq:hamilt}
\end{align}
Under Assumption \ref{ass:cost},  
the minimization in \eqref{eq:hamilt} is straightforward to analyze (and in general is related to the Fenchel transform of $c(\cdot)$). Moreover the value $y^*$ realizing the minimum is unique and the following mapping,
\begin{align}
    x \mapsto \beta_{opt}(x,I_t^{\bar \beta}) = y ^* = \argmin_{y \in [\bmin,\beta_0]}
\Big\{ c (y) + (r_I-x) \cdot y \cdot I_t^{\bar \beta} \Big\},
\end{align}
is Lipschitz in both arguments with a constant $L_H$ valid for all $x\in [0,r_I]$ and $I_t^{\bar \beta} \in [0,1]$. 
Since $I_t^{\bar \beta}$ is Lipschitz (in particular continuous) we obtain that 
 $\beta_{opt}(\Pi_t,I_t^{\bar \beta})$ is continuous with respect to time and thus $\Pi_t$ is a classical solution of \eqref{eq:hamilt}. It is a Lipschitz function of time with Lipschitz constant $L_\Pi \le c(\beta_0)+r_I \beta_0$.
\end{proof}

We now prove Theorem \ref{thrm:existence}. 

\begin{proof}
The proof consists of applying Schauder Theorem 
(see for example \cite[Theorem 1.C]{zeidler2012applied} or \cite[Theorem 18.20]{poznyak2008topics})
to the mapping $\Tc$, in order to prove that it has a fixed point. To this end, we define a subset $\Dc \subset \Bc$ consisting of Lipschitz functions with constant
$L_H (L_\Pi + \beta_0 + \gamma)$. Obviously $\Dc$ is a compact set in $L^2(0,T)$. Previous considerations show that for any ${\bar \beta}$, the corresponding  optimal individual choice
$\beta_{opt}(\Pi_t,I_t^{{\bar \beta}})$ is in $\Dc$. Moreover $\Dc$ is a compact subset of $L^2(0,T)$. For any
sequence ${\bar \beta}^n$ in $\Bc$ converging in $L^2$  to some ${\bar \beta}^{\infty}$ we have that on the one hand 
$I_t^{{\bar \beta}^n} \to I_t^{{\bar \beta}^{\infty}}$ for any $t$. Moreover the corresponding value functions $\Pi_t^n$ (which depend on ${\bar \beta}^n$ through $I_t^{{\bar \beta}^n}$)
will also converge pointwise.
This means that finally the optimal values
$\beta_{opt}(\Pi_t^n,I_t^{{\bar \beta}^n})$
will converge to the corresponding limits 
$\beta_{opt}(\Pi_t^{\infty},I_t^{{\bar \beta}^{\infty}})$.
By the dominated convergence theorem we obtain the $L^2$ convergence of 
$\Tc ({\bar \beta}^n)$ to 
$\Tc ({\bar \beta}^{\infty})$ which ends the proof, by application of the Schauder fixed point theorem. 
\end{proof}

\section{Conclusion and closing remarks}\label{sec:conclusion}

In this work, we consider, 
within both a SIR and SEIR models,
the question of the COVID-19 control measures as seen from the point of view of the individuals. We assume that each individual can choose to decrease his/her social interactions in order to slow down the spread of the virus. This means that the transmission rate, usually constant and exogenous in standard SIR models, is endogenous and time-dependent. Individuals can choose to lower their contact rate, which will allow them to minimize the likeliness of being infected, but this comes at a cost. The impact on the overall epidemic unfold from a single individual is negligible but the aggregating behavior of all individuals determine the epidemic evolution. This is formalized through a Mean Field approach.

We prove the existence of a Mean Field Nash equilibrium, \textit{i.e.}, a contact rate such that no individual has an interest in choosing a contact rate different from the overall societal contact rate. We then perform numerical simulations in order to find this equilibrium, which seems to be unique for the tested cases. The transmission rate of the disease induced by the equilibrium allows a clear improvement in the evolution of the epidemic, compared to the evolution with the initial transmission rate $\beta_0$, which corresponds to the transmission rate of the disease without any effort of the population. In particular, despite the selfishness of individuals (who only seek to minimize their own cost), their efforts make possible to reduce the number of people affected by the disease by $25\%$. In addition, the infection peak is less critical, which limits, and may even prevent, the saturation of the health care system and a corresponding decrease in the mortality rate of the virus.

However, the Nash equilibrium is not the best that can be achieved if compared with a situation where individuals are $100\%$ altruistic and only see the good of the society as a whole. To quantify this difference, we compute the societal-wide optimum to compare the two strategies. We observe that the divergence between the two strategies arrives both before and after the peak of the epidemic. More precisely, the Nash equilibrium allows, through premature efforts, to decrease the peak of infected, but the centralized epidemic control implies a more rapid decrease in the proportion infected after the peak, by maintaining intense efforts. As a consequence, there is a \textit{cost of anarchy}, meaning that the Nash equilibrium induces a higher cost than the societal optimum. While stopping early, at a point where the epidemic decreases but is not yet over, may be intuitively explained by the selfish nature of individuals, the fact that they decide collectively to begin efforts early than the societal optimum is a more intriguing feature. The latter fact is consistent with the experimental results documented in the remarkable work \cite{ghader2020observed}. 

Of course, any model is but an imperfect description of the reality and there are some limitations of our model too. First of all, as there is no consensus in the medical literature on the parameters of the epidemic, this impacts our model too. Indeed, the recent literature on the COVID-19 epidemic is abundant but discordant on the dynamics of the epidemic, in particular on the parameters $\gamma$ and $\mathcal R_0 := \beta_0 / \gamma$. Moreover, the appropriate estimation of the cost of effort, denoted by the function $c$, and its counterpart $r_I$ are still object of research, as they are not necessarily monetary or economic costs, but often costs related to health and social interactions. The modeling of these costs needs to be related to the concept of QALY/DALY, see \cite{zeckhauser1976now, anand1997disability, sassi2006calculating}. 

To account for the long latency phase of the disease induced by the COVID-19, we extend our reasoning from a SIR model to a SEIR. This choice can also be discussed since the disease induced by the COVID-19 has many other complex dynamics (for \textit{e.g.} large number of asymptomatic carriers, see \cite{ng2003ado,turinici_sars_2006, danchin_new_2020} for alternatives already used in coronavirus epidemics). 
Our model could also be extended to take into account other ways to control an epidemic, such as vaccination for example, even if no vaccine is known to date. Moreover we assume that all individuals are rational, identical, and that they have a perfect knowledge of the epidemic dynamics. The model could also be extended to a add some heterogeneity between individuals, since the epidemic does not affect individuals in the same way, and is more costly for those at risk. All these remain for future work.

The study of the cost of anarchy raises the question of incentives: what levers can healthcare authorities use to bring the Nash equilibrium closer to the societal optimal? Indeed, many countries have introduced fines, or even prison sentences, in the event of failure to comply with lockdown. Our model can give details in this direction, and invites a finer assessment of the costs related to the epidemic, but also to the lockdown. A model with centralized control also opens up the question of modeling the interaction (or not) between countries with different levels of epidemic dynamics.

\newgeometry{margin=1.1in}
{\footnotesize
\bibliographystyle{abbrv}
\bibliography{biblioglobale}}

\restoregeometry

\begin{appendix}

\newpage

\section{Additional numerical results}\label{app:fig_sensibility}

The figures listed and described in Section \ref{sec:sensibility} are grouped together in this appendix. The parameters are those described in Table \ref{tab:params}, except when other values are explicitly mentioned. 

\begin{figure}[!ht]
\includegraphics[scale=0.65]{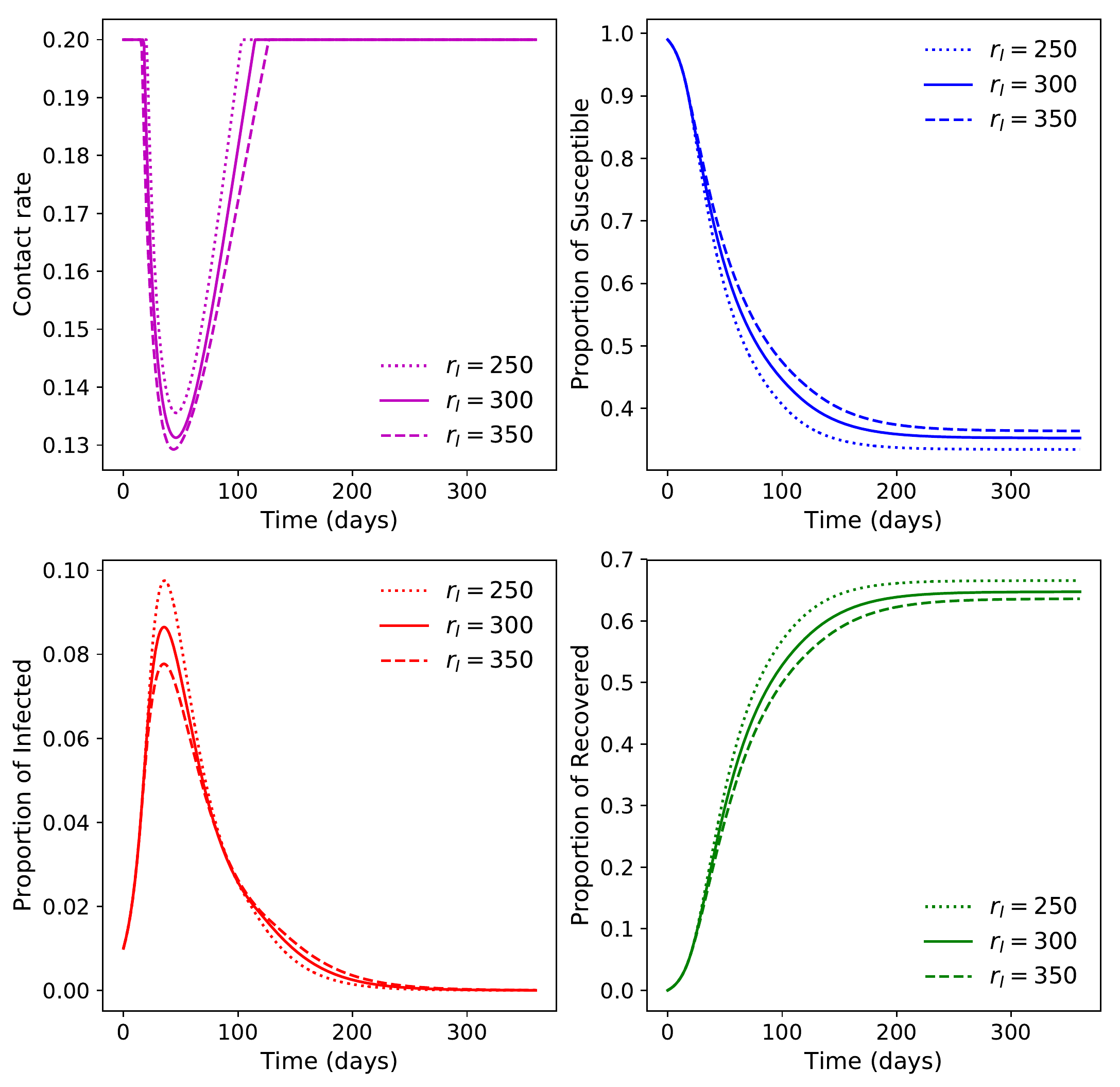}
\centering
\caption{Comparison of the Nash equilibrium contact rate and the evolution of Susceptible, Infected and Recover classes with three different values for $r_I$. Solid lines: $r_I = 300$. Dotted line: $r_I = 250$. Dashed lines: $r_I = 350$.} 
\label{fig:SIR_sensi_rI}
\end{figure}

\begin{figure}[!ht]
\includegraphics[scale=0.65]{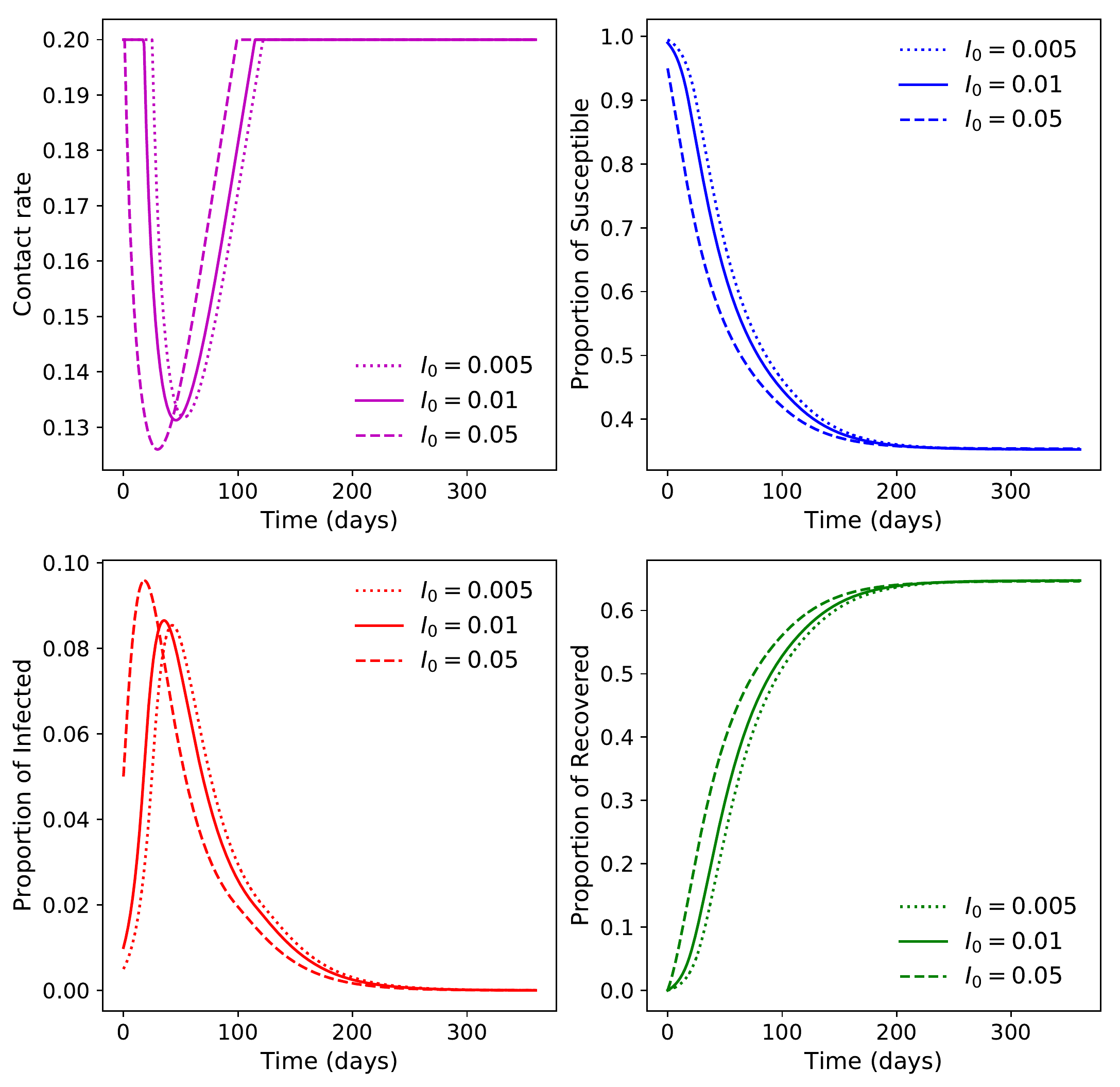}
\centering
\caption{Comparison of the Nash equilibrium contact rate and the evolution of Susceptible, Infected and Recover classes with three different values for $I_0$. Solid lines: $I_0 = 0.01$. Dotted line: $I_0 = 0.005$. Dashed lines: $I_0 = 0.05$.} 
\label{fig:SIR_sensi_i0}
\end{figure}

\begin{figure}[!ht]
\includegraphics[scale=0.65]{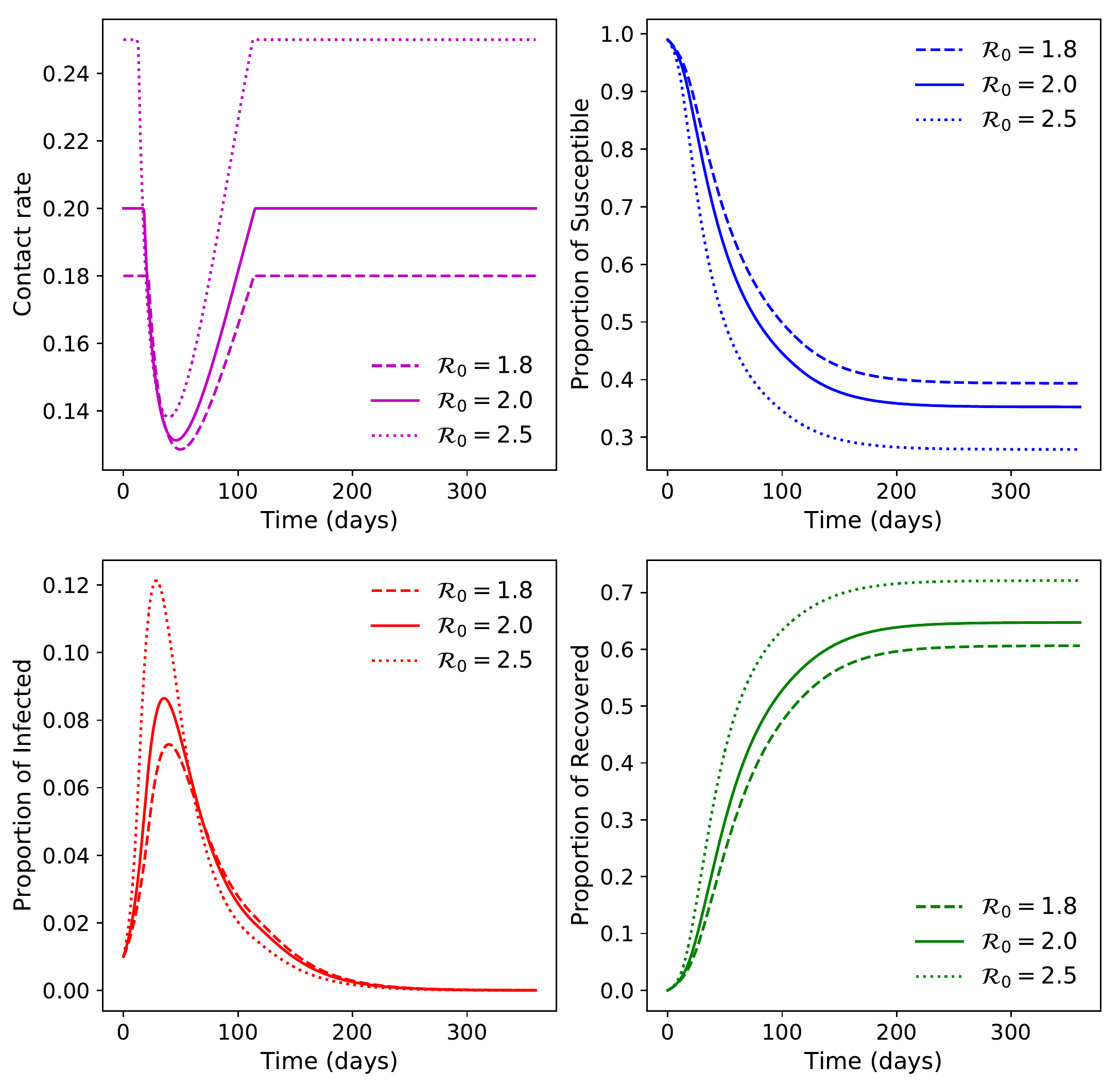}
\centering
\caption{Comparison of the Nash equilibrium contact rate and the evolution of Susceptible, Infected and Recover classes with three different values for $\mathcal R_0$. Dashed lines: $\mathcal R_0 = 1.8$. Solid lines: $\mathcal R_0 = 2.0$. Dotted line: $\mathcal R_0 = 2.5$.} 
\label{fig:SIR_sensi_beta_0}
\end{figure}

\end{appendix}

\end{document}